\documentclass[aps,prd,twocolumn,showpacs,amsmath,amssymb,nofootinbib]{revtex4-1}
\usepackage{blindtext}
\usepackage{graphicx}
\usepackage{booktabs}
\usepackage{color}
\usepackage{times}
\usepackage{bm}
\usepackage{multirow}
\usepackage{float}
\usepackage{url}
\usepackage{natbib}
\usepackage{tensor}
\usepackage{bbm}
\usepackage[colorlinks=true,citecolor=blue,urlcolor=blue,linkcolor=red]{hyperref}
\usepackage[normalem]{ulem}
\usepackage{verbatim}
\usepackage{lipsum} 
\usepackage{tikz,xcolor,hyperref}
\usepackage{ifxetex,ifluatex}
\usepackage{orcidlink}
\usepackage{lipsum} 
\usepackage{tikz,xcolor,hyperref}

\def\fm3{\;\text{fm}^{-3}}

\DeclareUnicodeCharacter{2212}{-}

\newcommand{\orcidauthorA}{0000-0001-9189-860X}
\newcommand{\orcidauthorB}{0000-0001-9849-3656}

\begin{document}
\title{Finite-Temperature Equations of State for Neutron-Star Mergers within the Quark Mean-Field Model
}

\author{Shuochong Han$^{1}$}
\altaffiliation{These authors contributed equally to this work.}
\author{Xiangdong Sun$^{2,1}$}
\altaffiliation{These authors contributed equally to this work.}
\author{Zhenyu Zhu{\orcidlink{\orcidauthorA}}$^{3}$}
\author{Ang Li{\orcidlink{\orcidauthorB}}$^{1}$}
\affiliation{\it 
$^{1}$Department of Astronomy, Xiamen University, Xiamen 361005, China; \textit{liang@xmu.edu.cn}; \\
$^{2}$Department of Physics and Astronomy, Qinghai University, Xining, 810016, China; \\
$^{3}$Center for Computational Relativity and Gravitation, Rochester Institute of Technology, Rochester, NY 14623, USA
}


\begin{abstract} 

We extend the quark mean-field (QMF) model to finite temperature and systematically investigate the thermodynamic properties of uniform nuclear matter, using the relativistic mean-field (RMF) model as a reference. 
For each model, we consider two parameter sets calibrated to the same selected nuclear-matter saturation properties but with different nucleon Dirac effective-mass ratios at saturation. 
We examine the effective nucleon mass and the composition of $\beta$-equilibrium matter, together with the free energy, energy density, pressure, entropy, and heat capacities of symmetric nuclear matter, pure neutron matter, and $\beta$-equilibrium matter. 
The QMF and RMF models exhibit similar qualitative trends, but their predictions increasingly differ at high density, where the effective masses have distinct density dependences. These differences are particularly evident in the thermal pressure and thermal index. 
For the two parameter sets considered in this work, we further provide compact analytical representations of the thermal index, $\Gamma_{\rm th}(\rho_{\rm b},T,\beta)$ over baryon densities of 0.08-1.0 $\rm fm^{-3}$, temperatures of 5-100 MeV, and isospin asymmetries of 0-1. 
The fits reproduce the numerical results with an accuracy better than $2\%$ and can be directly implemented in binary neutron-star merger simulations.
This study establishes a finite-temperature extension of the QMF framework and provides a basis for future assessments of how a quark-structure-based description of nuclear matter affects thermal support and post-merger dynamics in binary neutron-star mergers.

\end{abstract}

\pacs{
21.65.Cd, 
26.50.+x, 
26.60.-c, 
97.60.Jd, 
26.60.Kp 
     }

\maketitle 

\section{Introduction}

The equation of state (EoS) of cold dense matter is a fundamental ingredient in the description of neutron stars (NSs) and has been increasingly constrained by recent multimessenger observations. 
In particular, gravitational-wave (GW) observations of binary neutron-star 
(BNS) mergers, such as GW170817~\citep{2017PhRvL.119p1101A, 2018PhRvL.121p1101A}, together with X-ray measurements from the NICER mission~\citep{2019ApJ...887L..21R, 2019ApJ...887L..24M, 2021ApJ...918L..27R, 2021ApJ...918L..28M}, have provided important constraints on the mass-radius relation and the properties of cold NS matter~\citep[e.g.,][]{2019Univ....5..159L,2021ApJ...913...27L,2022NatRP...4..237Y,2023ApJ...943..163Z}. 
However, these observations primarily constrain the EoS of matter at relatively low temperatures, while the finite-temperature behavior of dense nuclear matter~\citep[e.g.,][]{2014PhRvC..89f5802C, 2021EPJA...57..329R, 2021PhRvC.103e5806K} remains less well studied. 

Thermal effects become particularly important in BNS mergers and core-collapse supernovae (CCSNe).
During these processes, shock heating and compression can raise the temperature of the remnant to several tens of MeV, with even higher temperatures occurring in localized hot regions~\citep{2023ApJ...952L..36F}. 
The resulting thermal pressure provides additional support against gravitational collapse and can therefore influence the structure, stability, and lifetime of the hypermassive neutron-star remnant~\citep{2020JHEAp..28...19L,2023PhRvD.108h3029R, 2025PhRvD.112l3008G}. 
Finite-temperature properties can also affect the neutrino emission and absorption in both BNS mergers and CCSNe, as well as the post-merger GW spectrum of BNS~\citep{2007PhRvL..99l1102O, 2010PhRvD..82h4043B, 2022MNRAS.516.4792R, 2023PhRvD.108f3032B, 2023ApJ...952L..36F, 2024PhRvD.110d3002R, 2025PhRvD.111h3009G, 2025PhRvD.112j3003Z, 2019PhRvC.100e5802S, 2020PhRvL.124i2701Y, 2020ApJ...894....4D, 2025ApJ...980...53E}.
A reliable description of hot, dense matter therefore requires an EoS that consistently incorporates density, temperature, and composition dependence.

A commonly used approximation in BNS merger simulations is to supplement a cold EoS with a phenomenological thermal component characterized by a constant thermal index $\Gamma_{\rm th}$~\citep[e.g.,][]{2010PhRvD..82h4043B,2020PhRvD.102d3006F,2023PhRvD.108h3029R}.
This index relates the thermal pressure, $P_{\rm th}$, to the thermal energy density, $\varepsilon_{\rm th}$, through $\Gamma_{\rm th}=1+P_{\rm th}/\varepsilon_{\rm th}$. 
Although this prescription is computationally convenient, microscopic finite-temperature EoS calculations have shown that the thermal response of nuclear matter is generally dependent on density, temperature, and composition, with its behavior closely connected to the quasiparticle spectrum and the density dependence of the nucleon effective mass~\citep{2014PhRvC..89f5802C,2015PhRvC..92b5801C,2021EPJA...57..329R}. 
Consequently, a constant $\Gamma_{\rm th}$ cannot fully capture the thermal behavior of matter over the broad range of thermodynamic conditions encountered in the post-merger remnant.
 These limitations motivate both the development of finite-temperature nuclear EoSs and the construction of compact representations of their thermal properties for numerical applications.
 
Conventional relativistic mean-field (RMF) models~\citep{2011PrPNP..66..519N} describe nucleons as effective degrees of freedom interacting through meson fields. Their in-medium properties are encoded in phenomenological meson–nucleon couplings. The quark mean-field (QMF) model~\citep{2000PhRvC..61d5205S,1998PhRvC..58.3749T} offers an alternative description that explicitly incorporates the internal quark structure of the nucleon~\citep{2019PhRvC..99b5804Z}.
In this framework, confined constituent quarks respond to the surrounding meson fields, and their dynamics determine the in-medium nucleon mass. The resulting effective scalar meson–nucleon coupling depends on the scalar field and consequently varies with the state of the medium~\citep{1988PhLB..200..235G}. 
This feature distinguishes QMF from conventional RMF models, leading to a different density dependence of effective nucleon masses and affecting the thermal responses of dense nuclear matter.

Previous applications of the QMF framework have primarily addressed cold nuclear matter, finite nuclei, and NS properties~\citep{2000PhRvC..61d5205S,1998PhRvC..58.3749T,2018PhRvC..97c5805Z,2018ApJ...862...98Z,2019PhRvC..99b5804Z,2018PrPNP.100..262G,2020JHEAp..28...19L,2023PhRvC.108b5809Z}. 
Some studies have examined low-temperature effects relevant for
NS cooling and thermal relaxation~\cite{2025PhRvD.112d3018T}.
Nevertheless, a systematic investigation of the thermodynamic properties of QMF matter over a broad range of temperature remains limited.
In this work, we extend the QMF model to finite temperature and systematically investigate the thermodynamic properties of nuclear matter up to temperatures of $T=50$ MeV, and we provide compact analytic parametrizations of the thermal index of the QMF model that are constructed and validated over the wider range $5\le T\le100$~MeV. 
We compare QMF and RMF parameter sets calibrated to the same selected nuclear saturation properties.
We first examine the in-medium nucleon mass and the composition of $\beta$-equilibrium matter, then compare the bulk thermodynamics of symmetric nuclear matter, pure neutron matter, and $\beta$-equilibrium matter. 
Finally, we briefly discuss the potential relevance of the resulting finite-temperature 
EoS to BNS merger simulations, particularly in connection with the thermal support of the post-merger remnant and the use of a density- and temperature-dependent $\Gamma_{\rm th}$.

The paper is organized as follows. Section II presents the QMF and RMF models, their finite-temperature thermodynamic formulation, and the parameter sets adopted in work. Section III examines the microscopic properties of finite-temperature matter, focusing on the scalar coupling, nucleon effective mass, and composition of $\beta$-equilibrium matter. Section IV compares the bulk thermodynamic state variables, including the free energy, energy density, pressure, entropy, and specific heats. Section V analyzes the thermal contributions and thermal index, presents compact analytic parametrizations for the QMF model, and discusses their relevance to NS mergers. Finally, Sec. VI summarizes our main results and outlines directions for future work.

\section{The QMF Model at Finite Temperature}
\label{sec:model}

We extend the QMF model to finite temperature and compare its predictions with those of the RMF model using parameter sets calibrated to the same selected nuclear-matter saturation properties. The underlying zero-temperature formulations have been described in e.g., Refs. ~\citep{2018PhRvC..97c5805Z,2018ApJ...862...98Z,2023PhRvC.108b5809Z} and are not repeated here. We instead summarize the model differences relevant to the present study and specify the thermal treatment and composition constraints. 

\subsection{Effective nucleon mass and comparison with RMF} 
\label{sec:iia}

The principal distinction between QMF and RMF is their treatment of the nucleon response to the scalar field~\citep{1998PhRvC..58.3749T}. In QMF, the in-medium nucleon Dirac effective mass, $m^*(\sigma)$, is obtained from the dynamics of confined constituent quarks, including center-of-mass, pionic, and gluonic corrections. The associated effective scalar coupling is therefore field dependent,
\begin{equation}
g_{\sigma N}(\sigma) = -\partial m^* / \partial \sigma
\end{equation}
In the RMF model considered here, the nucleon is treated as a structureless effective degree of freedom, with
\begin{equation}
m^*=m-g_{\sigma N}\sigma,
\end{equation}
where $m$ is the vacuum nucleon mass and $g_{\sigma N}$ is constant.  
In QMF, $g_{\sigma q}$, $g_{\omega q}$, and $g_{\rho q}$ denote the couplings of constituent quarks to the scalar-isoscalar $\sigma$, vector-isoscalar $\omega$, and vector-isovector $\rho$ mesons, respectively; the corresponding meson–nucleon couplings in RMF are $g_{\sigma N}$, $g_{\omega N}$, and $g_{\rho N}$. In both models, $g_2$ and $g_3$ are the coefficients of the cubic and quartic scalar self-interactions~\citep{BOGUTA1977413}, while $\Lambda_v$ controls the mixed $\omega$-$\rho$ interaction~\citep{2001PhRvL..86.5647H}. The mean-field equations, energy functionals, and coupling conventions follow Ref. \citep{2018ApJ...862...98Z}, with the scalar response evaluated consistently in each framework.

The finite-temperature extension retains the QMF quark ground-state construction at the self-consistent scalar field. Consequently, temperature modifies the internal nucleon structure indirectly through changes in the mean fields; explicit thermal excitation of confined quarks is not included.

For each model, we employ two parameter sets with effective-mass ratios at saturation $m_0^*/m$ = 0.77 and 0.80, denoted Sets 1 and 2, respectively. The remaining selected saturation properties are common to both sets and both models, as summarized in Table \ref{tab:NM properties}. The model parameters are listed in Table~\ref{tab:coupling}. 
This calibration provides a common zero-temperature reference while allowing different scalar responses and thermodynamic behavior away from saturation.

\begin{table}[t]
\caption{Nuclear-matter properties at saturation used to calibrate the QMF and RMF models. Sets 1 and 2 share the same saturation density $\rho_0$, binding energy per nucleon $E_0$, incompressibility $K_0$, symmetry energy $S_0$, and symmetry-energy slope $L_0$, but differ in the nucleon Dirac effective-mass ratio $m_0^*/m$.}
\label{tab:NM properties}
\begin{ruledtabular}
\setlength{\tabcolsep}{1pt}
\begin{tabular}{lcccccc}
&$\rho_0$ & $E_0$ & $K_0$ & $S_0$ & $L_0$ & $m_0^*/m$ \\
&(fm$^{-3}$) & (MeV) & (MeV) & (MeV) & (MeV) &  \\
\hline
Set~1/2&0.16 & $-16$ & 231 & 31 & 40 & 0.77/0.80 \\
\end{tabular}
\end{ruledtabular}
\end{table}

\begin{table}[t]
\caption{Coupling parameters of the QMF and RMF models calibrated to the saturation properties listed in Table \ref{tab:NM properties}. The parameters are defined in Sec.~\ref{sec:iia}. Sets 1 and 2 correspond to $m_0^*/m$ = 0.77 and 0.80, respectively.}
\label{tab:coupling}
\begin{ruledtabular}
\begin{tabular}{lcc}
Parameter & Set 1 (QMF) & Set 2 (QMF)\\
\hline
$g_{\sigma q}$ & 3.862036 & 3.688795\\
$g_{\omega q}$ & 2.917483 & 2.653600\\
$g_{\rho q}$ & 6.958808 & 1.784156\\
$g_2$ (fm$^{-1}$) & $-24.617959$ & $-28.206871$\\
$g_3$ & $-6.344246$ & $-4.935649$\\
$\Lambda_v$ & 0.769366 & 1.023745\\
\hline
Parameter & Set 1 (RMF) & Set 2 (RMF)\\
\hline
$g_{\sigma N}$ & 8.070138 & 7.684347\\
$g_{\omega N}$ & 8.752452 & 7.960801\\
$g_{\rho N}$ & 5.412945 & 5.373323\\
$g_2$ (fm$^{-1}$) & $-10.413850$ & $-11.954810$\\
$g_3$ & $-5.269550$ & 12.075241\\
$\Lambda_v$ & 0.096170 & 0.127968\\
\end{tabular}
\end{ruledtabular}
\end{table}

\begin{figure*}
{\centering
  \includegraphics[width=0.47\textwidth]{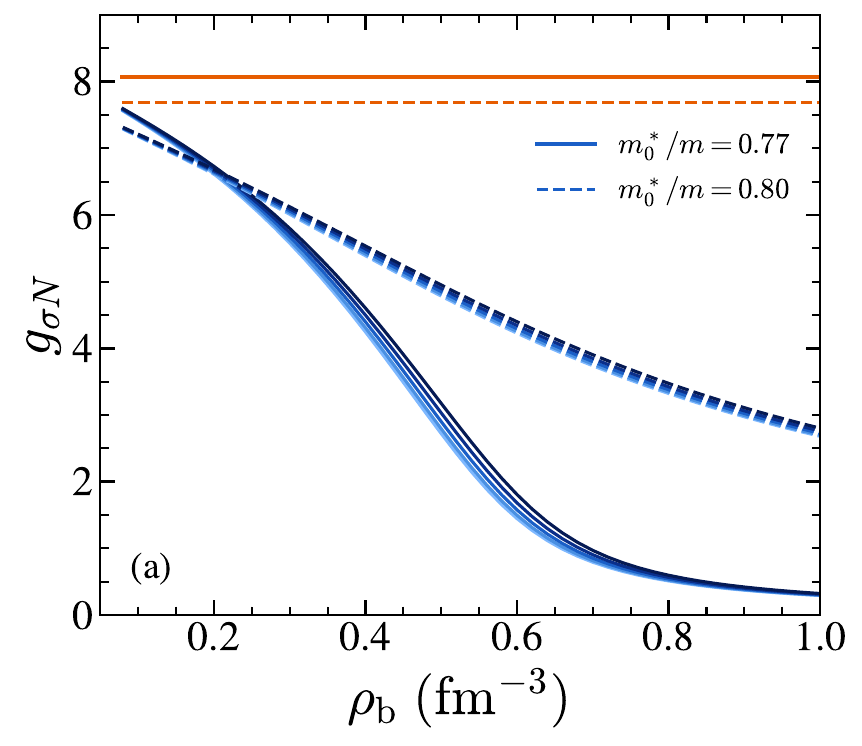}
  \includegraphics[width=0.49\textwidth]{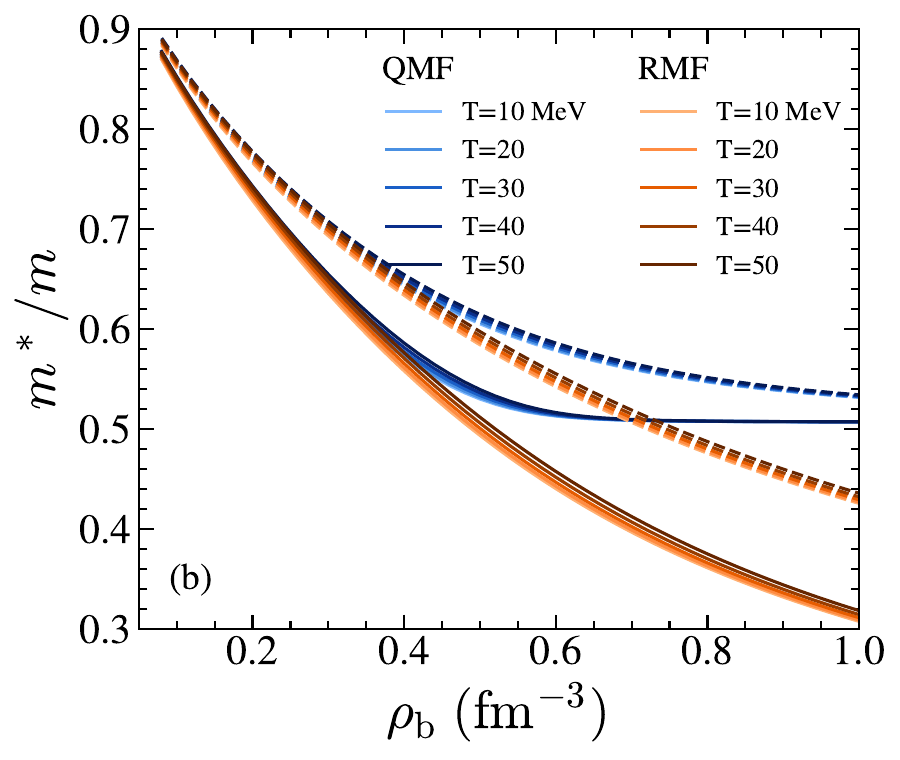}
  \caption{ 
  Effective scalar coupling $g_{\sigma N}$ (left) and nucleon Dirac effective-mass ratio $m^*/m$ (right) as functions of baryon density $\rho_{\rm b}$ in symmetric nuclear matter for the QMF (blue) and RMF (orange) models. Solid and dashed curves correspond to Sets 1 and 2, with $m_0^*/m$ = 0.77 and 0.80, respectively. Increasing color intensity denotes $T$ = 10, 20, 30, 40, and 50 MeV.}
  \label{fig:gsig&meff}
  }\vspace{-0.1cm}
\end{figure*}

\subsection{Finite-temperature formulation and matter composition}

For uniform matter at temperature $T$, the nucleon occupation probabilities are
\begin{equation}
n_i(p) = 1 / \{ \exp[(E^*(p) - \nu_i)/T] + 1 \},
\end{equation}
where $i = n, p$, $E^*(p) = [p^2 + m^{*2}]^{1/2}$, and $\nu_i = \mu_i - V_i$ is the effective chemical potential after subtracting the vector mean-field contribution $V_i$. 

Neglecting thermal antinucleons, the neutron and proton densities and the total scalar density are
\begin{eqnarray}
\rho_i = \frac{1}{\pi^2} \int_0^\infty p^2 n_i(p)\, dp , \\
\rho_s = \frac{1}{\pi^2} \sum_{i=n,p} \int_0^\infty
p^2 \frac{m^*}{E^*(p)} n_i(p)\, dp.
\end{eqnarray}
The total baryon density and isospin asymmetry are defined as
\begin{eqnarray}
\rho_{\rm b}=\rho_n+\rho_p, \\
\beta=\frac{\rho_n-\rho_p}{\rho_{\rm b}}.
\end{eqnarray}
At each specified temperature, baryon density, and composition, the chemical potentials and meson fields are determined self-consistently. The finite-temperature mean-field equations retain the structure of the zero-temperature equations, with the source densities evaluated using the thermal distributions. In particular, the QMF scalar source contains $-\partial m^* / \partial \sigma$, while the corresponding RMF source contains the constant coupling $g_{\sigma N}$.

\begin{figure*}
{\centering
\includegraphics[width=0.9\textwidth]{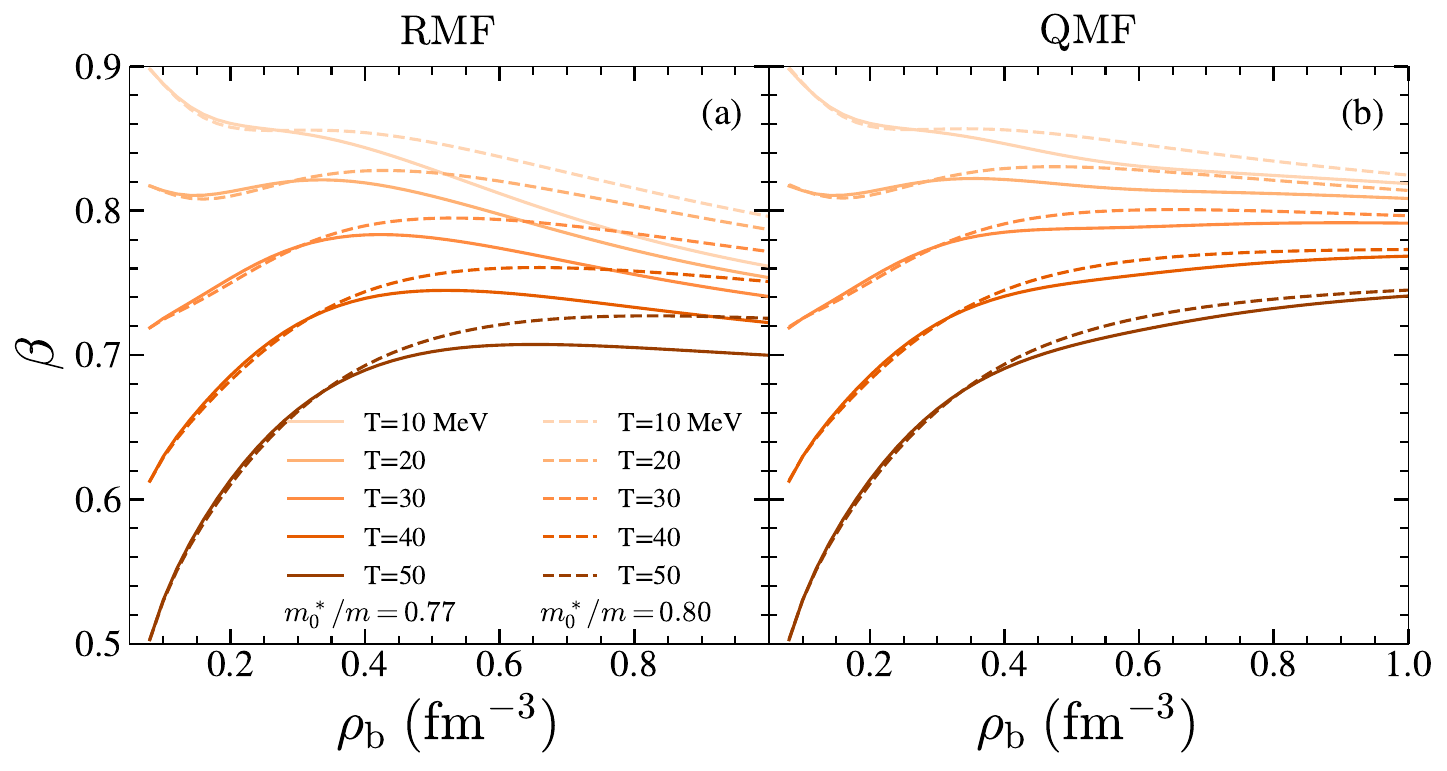}
  \caption{
  Isospin asymmetry $\beta$ as a function of baryon density $\rho_{\rm b}$ for charge-neutral, neutrino-transparent matter in $\beta$ equilibrium, calculated with the RMF (left) and QMF (right) models. Solid and dashed curves correspond to $m_0^*/m$ = 0.77 and 0.80, respectively. Increasing color intensity denotes $T$ = 10, 20, 30, 40, and 50 MeV.
  }
  \label{fig:beta}
  }\vspace{-0.1cm}
\end{figure*}

The nucleonic entropy density is
\begin{equation}
s_N = -\frac{1}{\pi^2} \sum_{i=n,p} \int_0^\infty p^2
\{
n_i \ln n_i + (1-n_i)\ln(1-n_i)
\}\, dp .
\end{equation}
The energy density is evaluated from the same mean-field energy functional, retaining the nonlinear meson contributions. The free-energy density and pressure then follow from
\begin{equation}
f_N = \varepsilon_N - T s_N, \
P_N = \mu_n \rho_n + \mu_p \rho_p - f_N.
\end{equation}
These relations are evaluated using the self-consistent fields, ensuring consistency between the energy density, entropy, chemical potentials, and pressure. The free energy per baryon and entropy per baryon are $f_N/\rho_{\rm b}$ and $s_N/\rho_{\rm b}$, respectively.

We consider symmetric nuclear matter, pure neutron matter, and charge-neutral, neutrino-transparent matter in $\beta$ equilibrium. For the first two systems, the composition is fixed at $\beta$ = 0 and $\beta$ = 1, respectively, and only nucleonic contributions are included.

For $\beta$-equilibrium matter, electrons, muons, and their antiparticles are treated as free relativistic Fermi gases. Their net densities are
\begin{eqnarray}
\rho_e = \rho_{e^-} - \rho_{e^+},\\
\rho_\mu = \rho_{\mu^-} - \rho_{\mu^+}.
\end{eqnarray}
The composition is determined by
chemical equilibrium, 
\begin{eqnarray}
\mu_n - \mu_p = \mu_e = \mu_\mu, 
\end{eqnarray}
and charge neutrality, 
\begin{eqnarray}
\rho_p = \rho_e + \rho_\mu.
\end{eqnarray}
These conditions are solved together with the meson-field equations at each $\rho_{\rm b}$ and $T$. The total thermodynamic quantities include both nucleonic and leptonic contributions:
\begin{eqnarray}
\varepsilon = \varepsilon_N + \varepsilon_e + \varepsilon_\mu, \\
P = P_N + P_e + P_\mu, \\
s = s_N + s_e + s_\mu.
\end{eqnarray}
For fixed-composition matter, thermodynamic derivatives are taken at fixed composition. For $\beta$-equilibrium matter, the composition is recalculated along the equilibrium sequence.

\section{Microscopic Properties at Finite Temperature}

In this section, we examine the effects of finite temperature on nucleon effective masses and particle fractions in dense matter, comparing the predictions of the RMF and QMF models. 
All results presented here and in subsequent sections are obtained for uniform nuclear matter; nonuniform phases are not considered. A unified QMF EoS incorporating the nonuniform low-density regime has been developed in Ref.~\cite{2025PhRvD.112d3018T}.

\subsection{Effective Mass ($m^*/m$)}

Figure \ref{fig:gsig&meff} compares the effective scalar coupling  $g_{\sigma N}$ and the nucleon effective-mass ratio $m^*/m$  at $T$ = 10–50 MeV. In RMF, $g_{\sigma N}$ is constant, whereas the QMF effective coupling decreases with density and increases slightly with temperature. This reduction in the QMF scalar coupling is accompanied by a progressively weaker density dependence of the effective mass.

The two models yield similar effective masses at low densities but differ substantially at high densities~\citep{2023PhRvC.108b5809Z, 2025PhRvD.112d3018T}. The RMF effective mass continues to decrease throughout the density range shown, while the QMF effective mass decreases more slowly and approaches a nearly constant value, particularly for Set 1. Consequently, QMF predicts larger effective masses at high densities for both parameter sets. The difference between the two sets is also smaller in QMF than in RMF in this regime.

In both models, the effective mass increases slightly with temperature, but its temperature dependence remains weak compared with its density dependence. The principal microscopic distinction is therefore the high-density scalar response, which underlies the differences in thermal properties discussed in the following sections.

\begin{figure*}
{\centering
\includegraphics[width=0.7\textwidth]{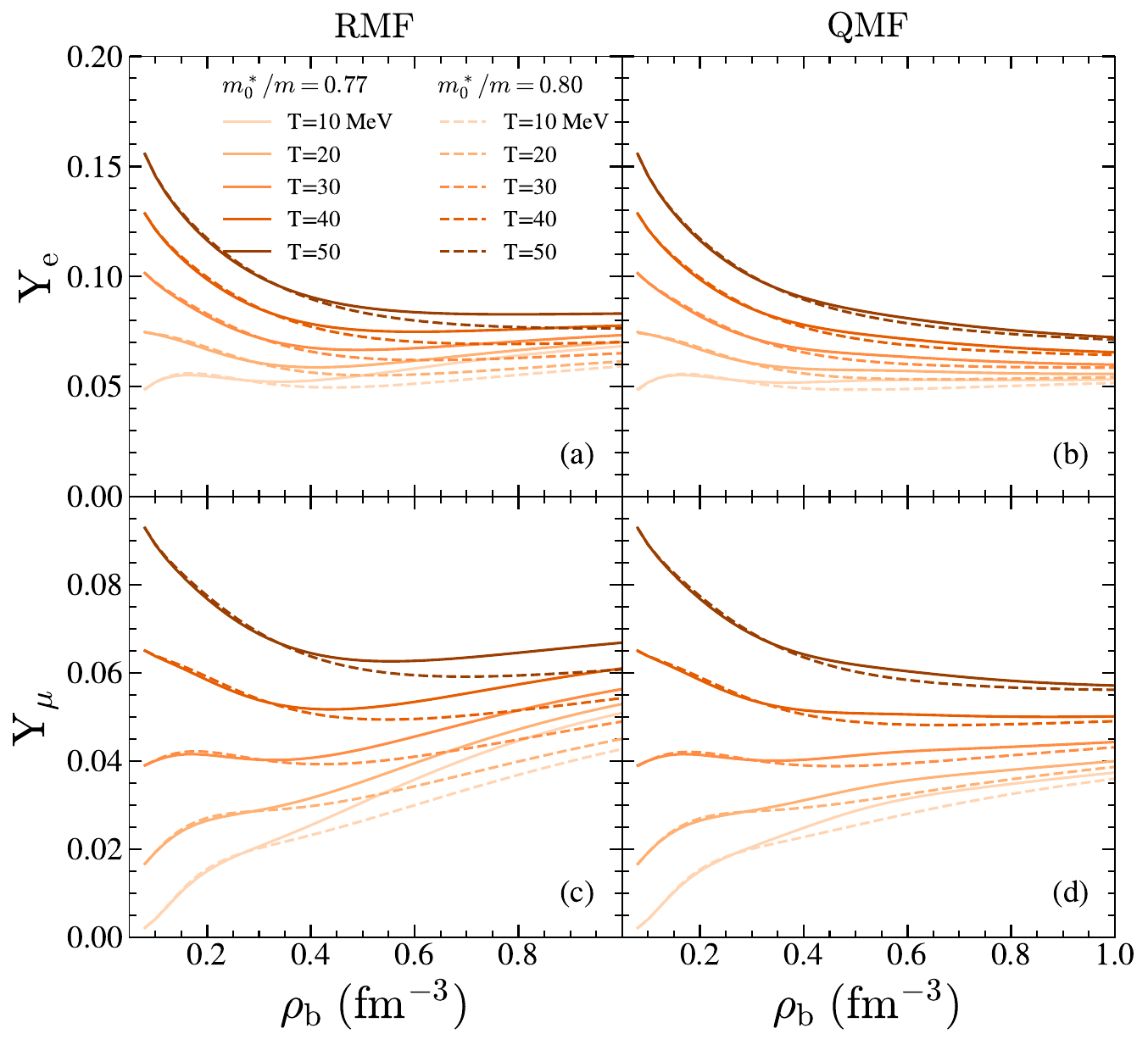}
  \caption{Net electron fraction $Y_e = \rho_e/\rho_{\rm b}$ (upper panels) and net muon fraction $Y_\mu = \rho_{\mu}/\rho_{\rm b}$ (lower panels) in charge-neutral $\beta$-equilibrium matter for the RMF (left) and QMF (right) models. The densities $\rho_e$ and $\rho_{\mu}$ denote particle densities minus the corresponding antiparticle densities.  Line styles and temperature colors follow Fig. \ref{fig:beta}.}
  \label{fig:lepton}
  }\vspace{-0.1cm}
\end{figure*}

\begin{figure*}
{\centering
\includegraphics[width=0.7\textwidth]{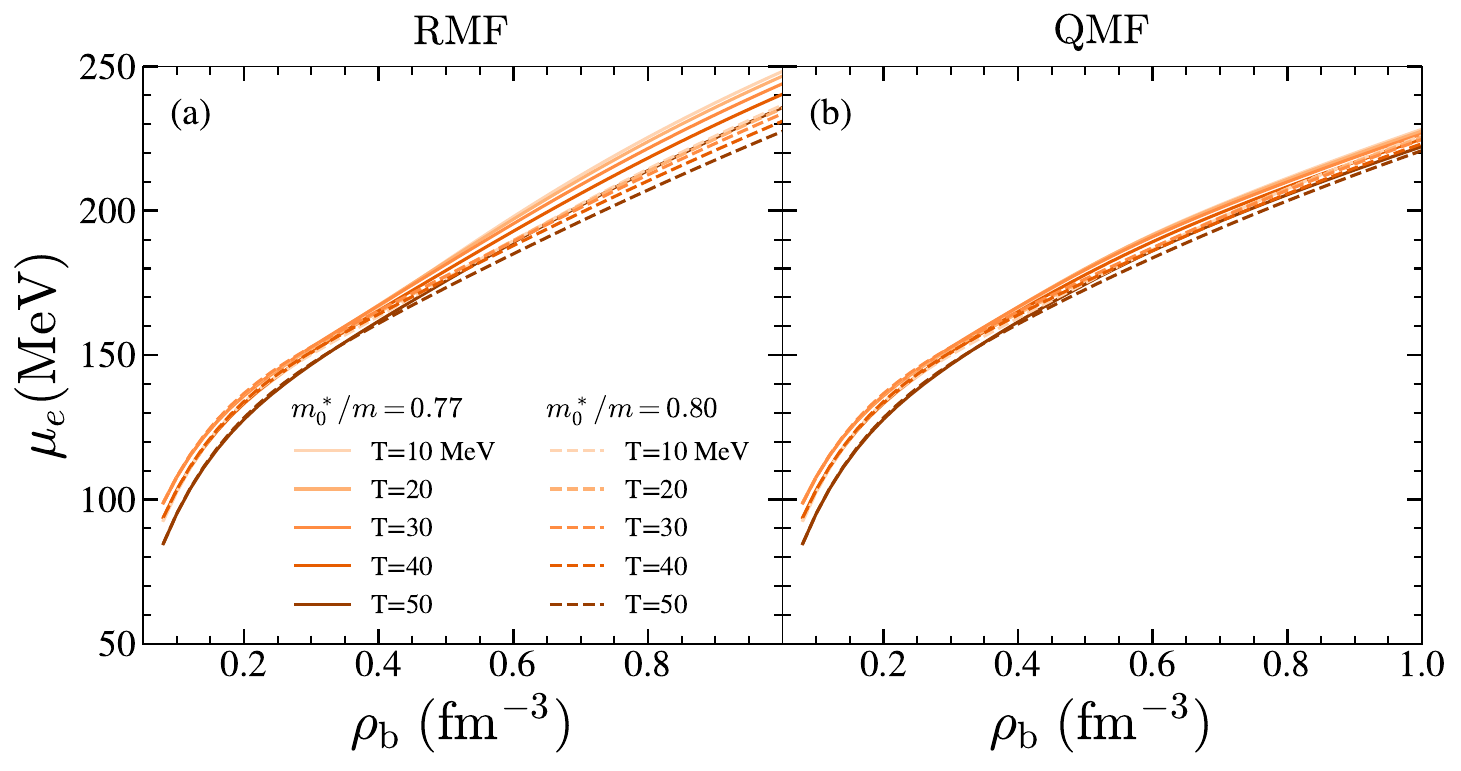}
  \caption{Electron chemical potential as a function of baryon density. Under the adopted equilibrium conditions, the electron and muon chemical potentials are equal, $\mu_e=\mu_\mu$; the plotted quantity therefore represents their common chemical potential. Line styles and temperature colors follow Fig. \ref{fig:beta}.
  }\label{fig:mu}
  }\vspace{-0.1cm}
\end{figure*}

\subsection{Particle Fractions ($Y_e, Y_\mu$)}

Figure \ref{fig:beta} shows the isospin asymmetry $\beta$ of charge-neutral matter in $\beta$ equilibrium. Both models exhibit a pronounced temperature dependence at low densities: increasing temperature lowers $\beta$ and therefore increases the proton fraction. At higher densities, the temperature dependence becomes less pronounced, although appreciable differences remain over the range shown.
The model dependence becomes apparent above approximately $\rho_{\rm b}$ = 0.3–0.4 $\rm{fm^{-3}}$. QMF predicts a larger isospin asymmetry than RMF and hence a smaller proton fraction. In both models, Set 2 generally gives a larger $\beta$ than Set 1 at high densities, but the separation between the parameter sets is smaller in QMF. These composition differences reflect the distinct neutron–proton chemical-potential differences generated by the self-consistent mean fields.

\begin{figure*}
{\centering
\includegraphics[width=0.95\textwidth]{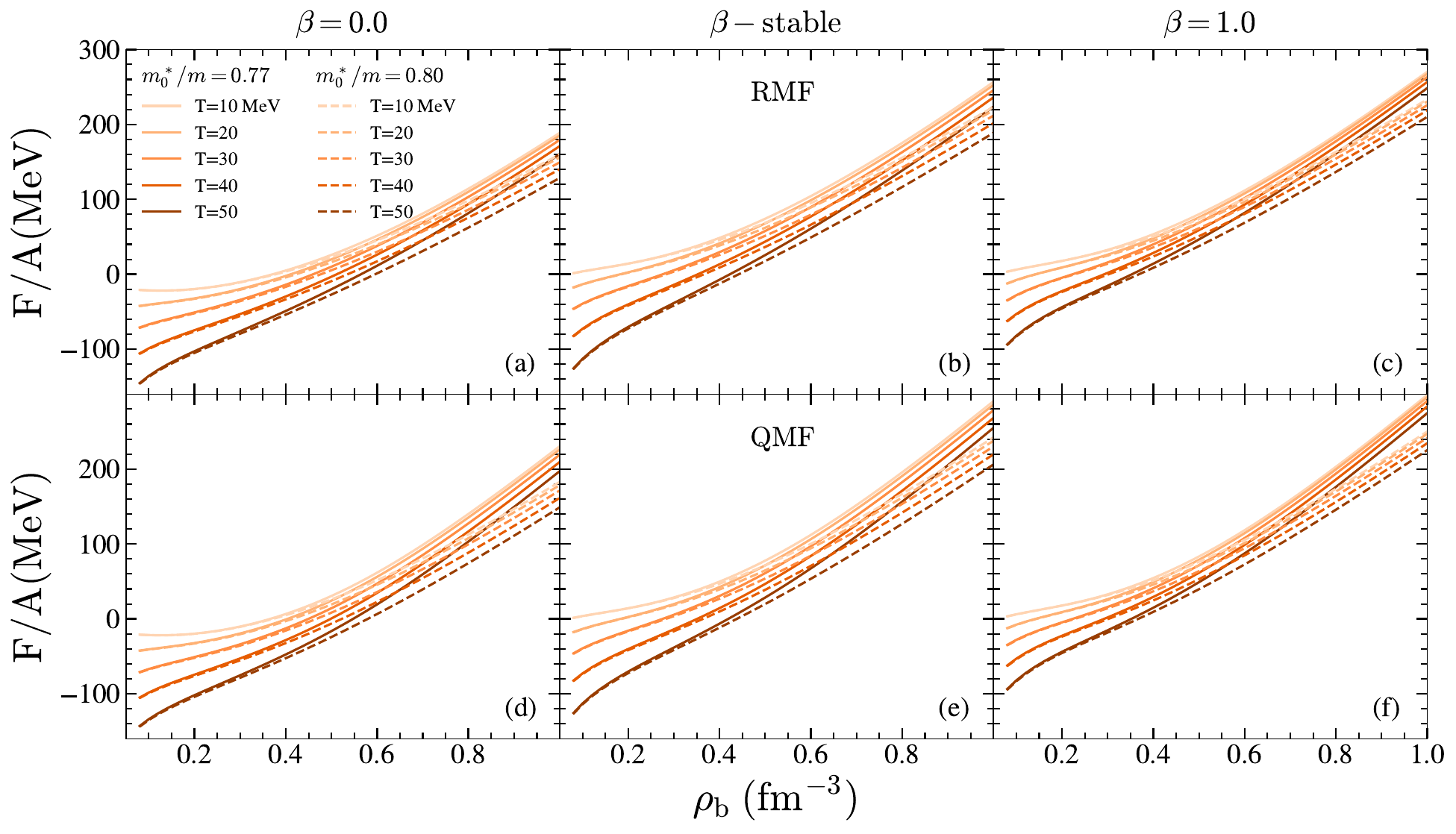}
  \caption{
  Free energy per baryon as a function of baryon density in the RMF (upper row) and QMF (lower row) models. The nucleon rest-mass contribution is subtracted. The left, middle, and right columns show symmetric nuclear matter ($\beta$ = 0), $\beta$-equilibrium matter, and pure neutron matter ($\beta$ = 1), respectively. Solid and dashed curves denote Sets 1 and 2, corresponding to effective-mass ratios at saturation $m_0^*/m$ = 0.77 and 0.80. Progressively darker colors indicate $T$ = 10, 20, 30, 40, and 50 MeV.}
  \label{fig:FA}
  }\vspace{-0.1cm}
\end{figure*}

Figure \ref{fig:lepton} presents the net electron and muon fractions. Their combined contribution equals the proton fraction through charge neutrality. Increasing temperature enhances the lepton fractions most strongly at low densities, while the low-temperature onset of the muon population becomes smoother as thermal occupation increases. At high densities, QMF generally predicts smaller electron and muon fractions than RMF, consistent with its larger isospin asymmetry.
In NSs, the contribution of leptons is non-negligible. 
In particular, the fractions of electrons and muons impact the URCA reaction rates, which determine the bulk viscosity and other dissipative processes~\citep{2025PhRvD.112d3018T}. These effects from the high-density and low-temperature ($\lesssim 15$\,MeV) core can be important for the GW evolution during the BNS inspiral and post-merger phases~\citep{2018PhRvL.120d1101A, 2018PhRvC..98f5806A, 2022MNRAS.509.1096M, 2024ApJ...967L..14M, 2024PhRvL.132u1001E, 2025PhRvD.111d4074C, 2025PhRvL.134g1402C, 2026arXiv260323904H, 2026arXiv260906374Z,
2023PhRvD.108j3037R, 2024NatAs...8.1277R}.

The corresponding lepton chemical potentials are shown in Fig. \ref{fig:mu}. Under the equilibrium conditions adopted here, electrons and muons share a common chemical potential, $\mu_e = \mu_\mu$. This chemical potential increases with density and is generally lower in QMF than in RMF at high densities. Its sensitivity to the parameter set is also weaker in QMF, consistent with the smaller separation between the composition curves in Figs. \ref{fig:beta}  and \ref{fig:lepton}.

\section{Thermodynamic State Variables}

In this section, we compare the thermodynamic properties of the finite-temperature QMF and RMF EoSs. We consider symmetric nuclear matter, pure neutron matter, and $\beta$-equilibrium matter, examining their dependence on baryon density, temperature, and the effective-mass ratio at saturation. Symmetric nuclear matter and pure neutron matter include only nucleonic contributions, whereas $\beta$-equilibrium matter also includes leptonic contributions. 

\subsection{Free Energy ($F/A$) \& Pressure ($P$)}

\begin{figure*}
{\centering
\includegraphics[width=0.95\textwidth]{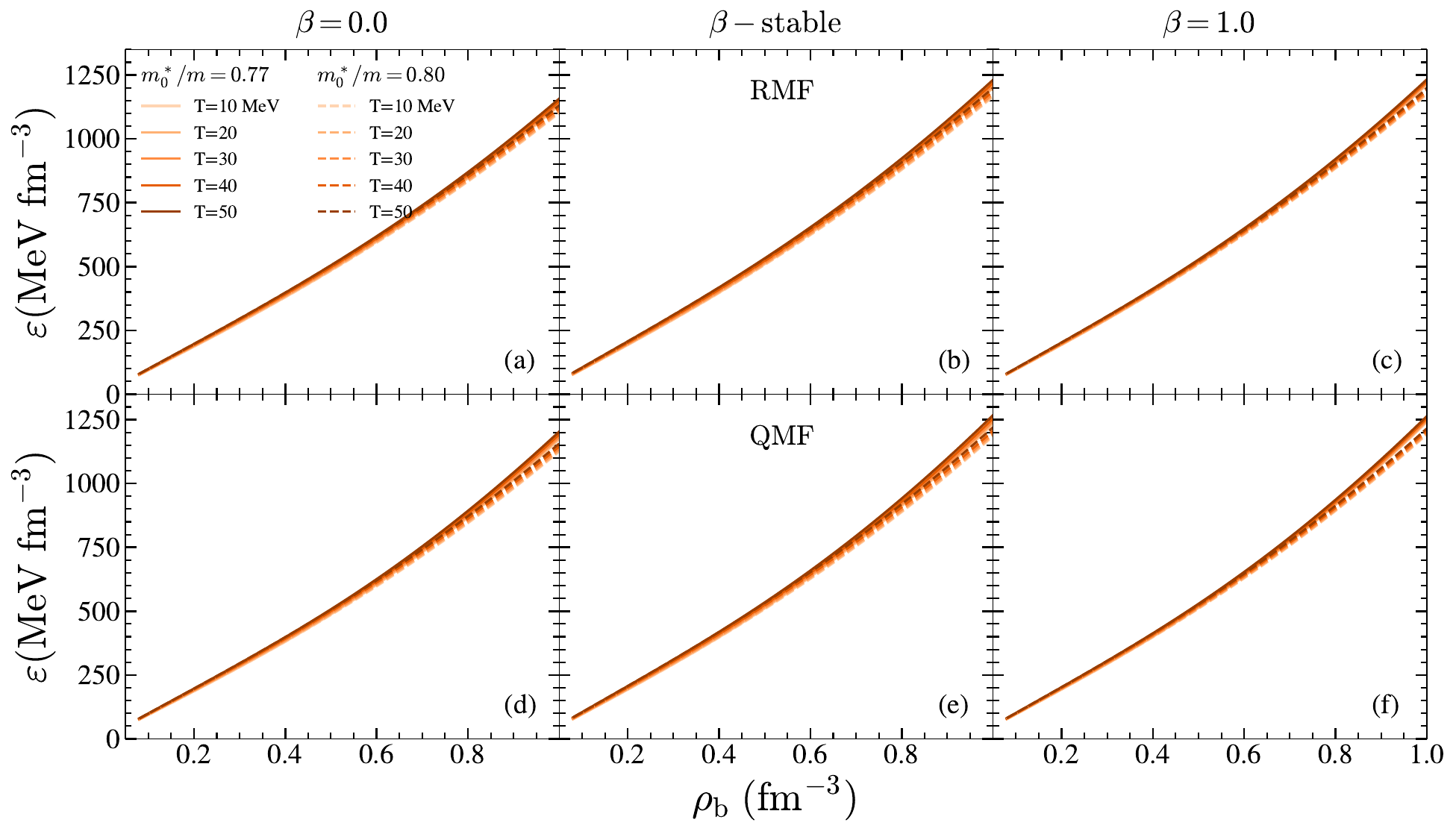}
  \caption{
  Energy density as a function of baryon density in the RMF (upper row) and QMF (lower row) models. Columns show symmetric nuclear matter, $\beta$-equilibrium matter, and pure neutron matter from left to right. Solid and dashed curves denote Sets 1 and 2, respectively; progressively darker colors indicate temperatures from 10 to 50 MeV in steps of 10 MeV.}
  \label{fig:epsilon}
  }\vspace{-0.1cm}
\end{figure*}

\begin{figure*}
{\centering
\includegraphics[width=0.95\textwidth]{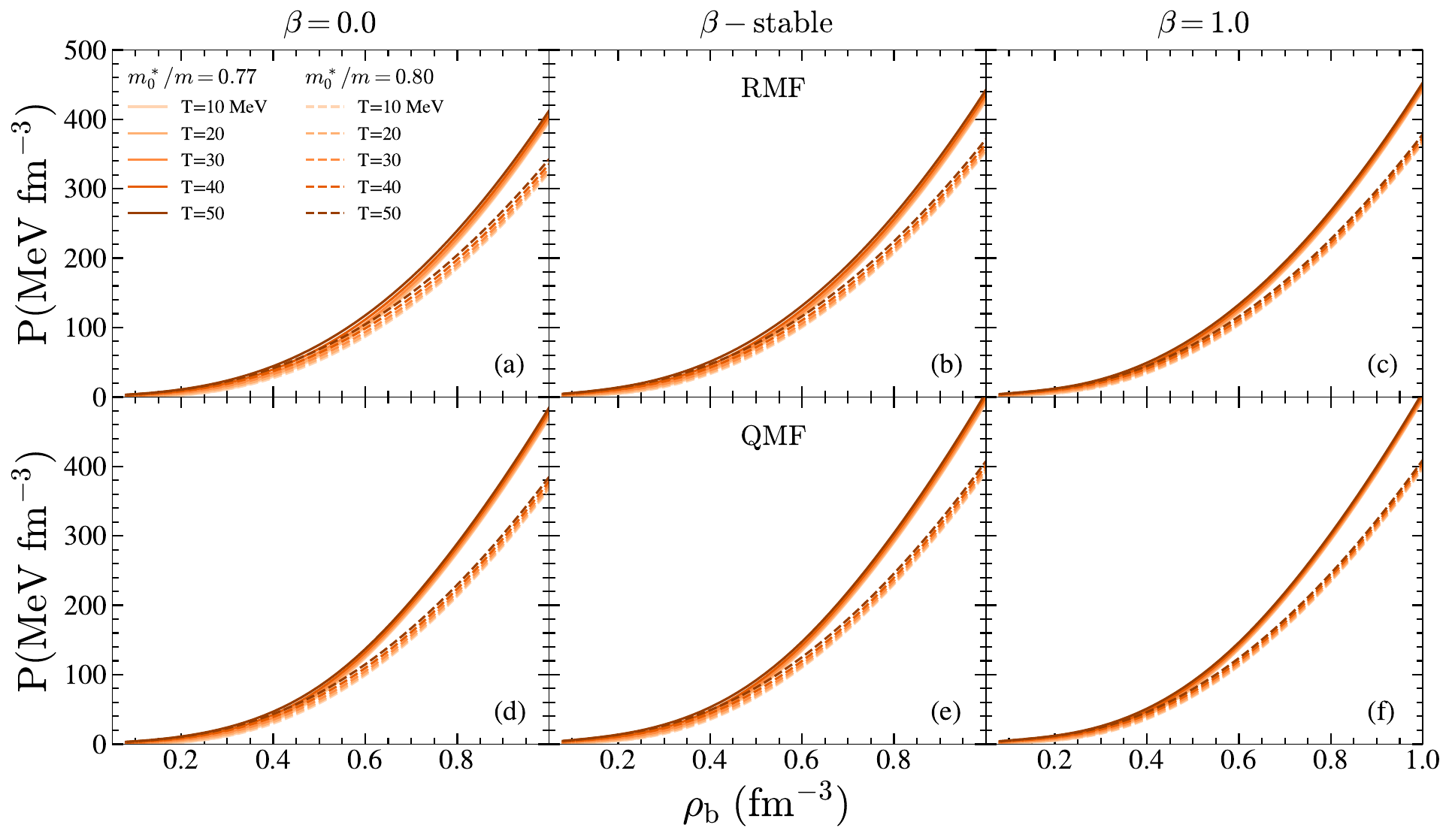}
  \caption{
  Pressure $P$ as a function of baryon density $\rho_{\rm b}$ in the RMF (upper row) and QMF (lower row) models. Columns show symmetric nuclear matter, $\beta$-equilibrium matter, and pure neutron matter from left to right. Solid and dashed curves denote Sets 1 and 2, respectively; progressively darker colors indicate temperatures from 10 to 50 MeV in steps of 10 MeV.}
  \label{fig:press}
  }\vspace{-0.1cm}
\end{figure*}

Figures \ref{fig:FA}–\ref{fig:press} show the free energy per baryon, energy density, and pressure, respectively. The two models exhibit similar qualitative trends, but their predictions increasingly differ at high density despite being calibrated to the same selected saturation properties.

As shown in Fig. \ref{fig:FA}, the free energy per baryon decreases with increasing temperature, consistent with the thermodynamic relation $F/A = E/A - TS$, where $S$ denotes the entropy per baryon. At low temperatures, symmetric nuclear matter exhibits a minimum associated with nuclear binding; this feature becomes less pronounced as the temperature increases. Increasing the neutron excess generally raises the nucleonic free energy. Both models also predict lower free energies for Set 2, which has the larger effective-mass ratio at saturation. The QMF–RMF differences depend on density and composition and become most apparent in the high-density region.

\begin{figure*}
{\centering
\includegraphics[width=0.95\textwidth]{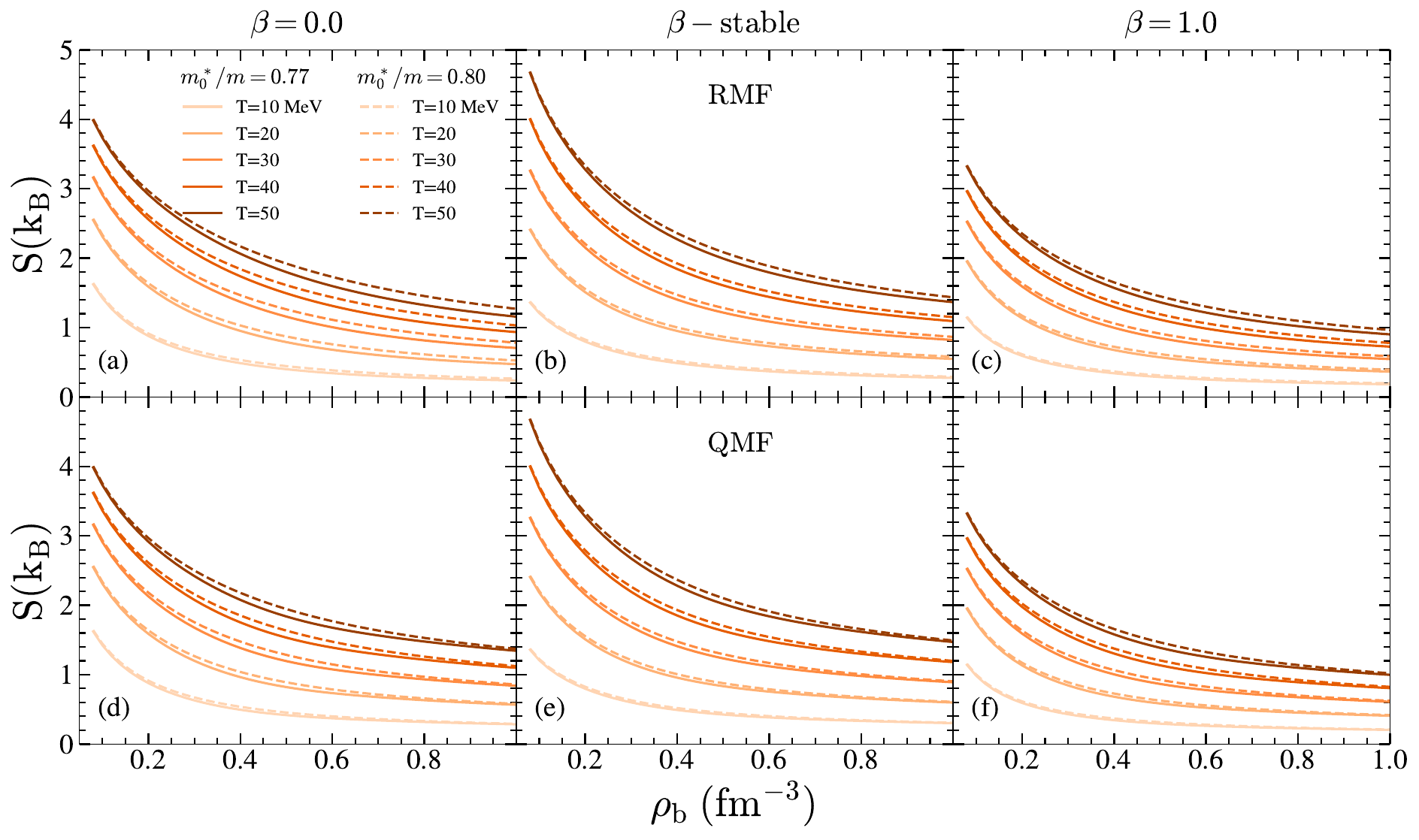}
  \caption{Entropy per baryon $S = s/\rho_{\rm b}$, in units of $k_{\rm B}$, as a function of baryon density $\rho_{\rm b}$ in the RMF (upper row) and QMF (lower row) models. Columns show symmetric nuclear matter, $\beta$-equilibrium matter, and pure neutron matter from left to right. Solid and dashed curves denote Sets 1 and 2, respectively; progressively darker colors indicate temperatures from 10 to 50 MeV in steps of 10 MeV.}
  \label{fig:Entropy}
  }\vspace{-0.1cm}
\end{figure*}

The energy density and pressure increase strongly with baryon density, as shown in Figs. \ref{fig:epsilon} and \ref{fig:press}. Their temperature dependence is comparatively modest on the scales displayed, particularly at high density, where the cold contributions dominate. For the parameter sets considered here, QMF generally predicts higher pressures than RMF at the same high baryon density~\cite{2023PhRvC.108b5809Z, 2025PhRvD.112d3018T}. In both models, Set 1 yields a higher high-density pressure than Set 2.
These differences reflect the distinct self-consistent mean fields and density dependences of the nucleon effective mass in the two models. The thermal contributions, which are less apparent in the total quantities, are isolated in Sec. V.

\subsection{Entropy ($S$)}

Figure \ref{fig:Entropy} shows the entropy per baryon, $S = s/\rho_{\rm b}$. At fixed temperature, $S$ decreases with increasing density over the range shown; at fixed density, it increases with temperature. Symmetric nuclear matter has a larger entropy than pure neutron matter, while $\beta$-equilibrium matter exhibits the largest entropy among the three systems considered, with leptons providing an additional contribution.
The QMF model generally predicts slightly higher entropies than RMF at high density. Within each model, Set 2 also yields higher entropies than Set 1. 

These trends are consistent with the role of the Landau effective mass in determining the density of quasiparticle states near the Fermi surface.
In the strongly degenerate limit, Landau Fermi-liquid theory gives $s=2T\sum_ia_i\rho_i$, where $a_i=(\pi^2/2)m^*_{L_i}/p^2_{\rm F_i}$ is the level-density parameter of species $i$~ \cite{2015PhRvC..92b5801C}.
For the momentum-independent mean fields used here, the nucleon Landau mass is $m^*_{L_i}=\sqrt{m^{\ast 2}_i + p_{\rm F_i}^2}$.
Thus, at fixed density, the entropy per particle increases approximately linearly with temperature. 
The results for $m^*_0/m=0.80$ are higher than those for $m^*_0/m=0.77$, since the effective mass for $m^*_0/m=0.80$ is larger in the high-density region.
This explains the increase of $S$ with temperature and effective mass, as well as the systematically larger entropy predicted by QMF when it yields a larger $m^*$ at high density.
Since $p_{\rm F_i}\propto\rho_i^{1/3}$, the entropy decreases monotonically with increasing density. 
The isospin dependence can be understood in the similar way.
SNM has smaller Fermi momenta than PNM at the same baryon density, resulting in a larger entropy.

\begin{figure*}
{\centering
\includegraphics[width=0.9\textwidth]{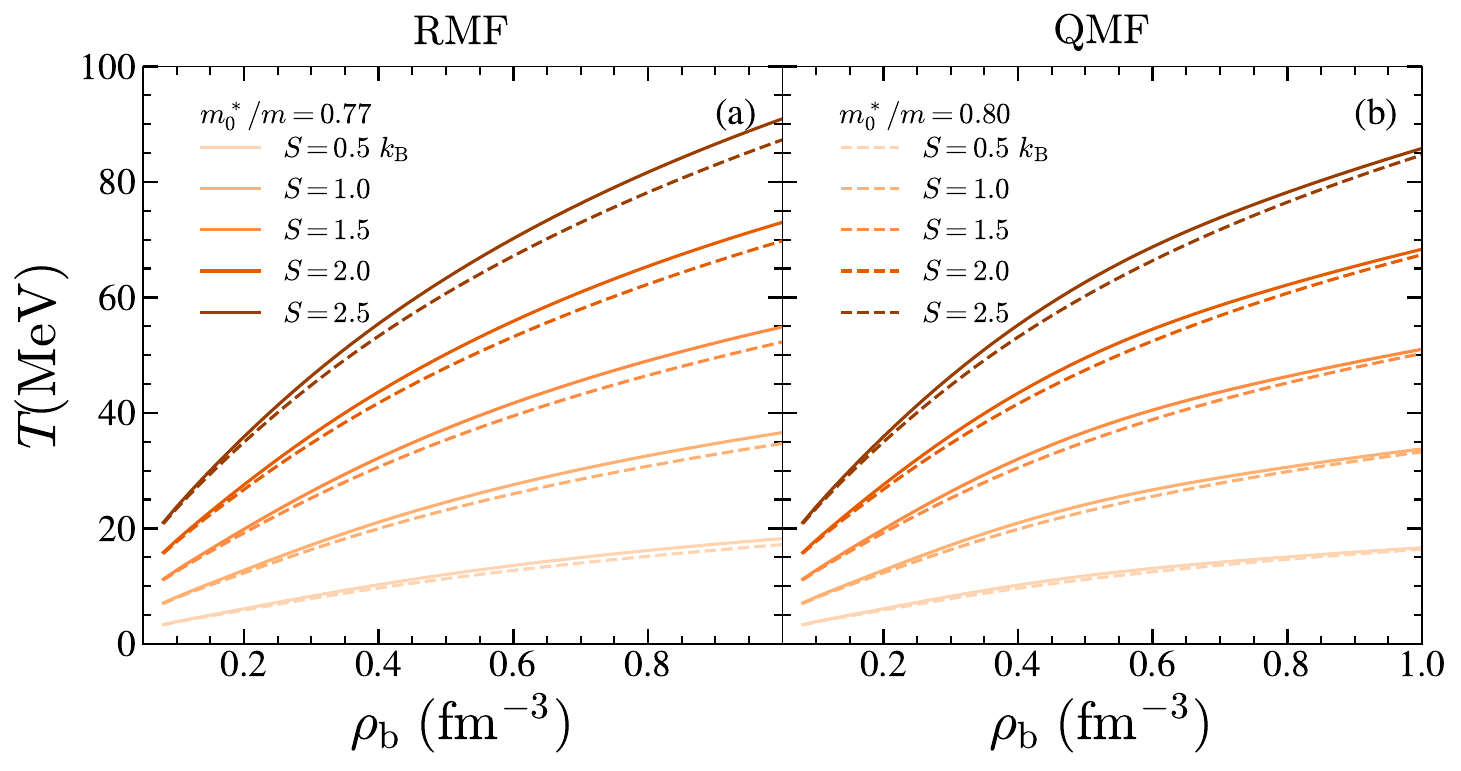}
  \caption{
  Isentropic contours in the baryon-density–temperature plane for $\beta$-equilibrium matter in the RMF (left panel) and QMF (right panel) models. Progressively darker colors correspond to entropies per baryon $S$ = 0.5, 1.0, 1.5, 2.0, and 2.5 $k_{\rm B}$. Solid and dashed curves denote Sets 1 and 2, with $m_0^*/m$ = 0.77 and 0.80, respectively.}
  \label{fig:TS}
  }\vspace{-0.1cm}
\end{figure*}

Figure \ref{fig:TS} presents isentropic contours for $\beta$-equilibrium matter in the $\rho_{\rm b}$-$T$ plane. The temperature increases with density along each contour, and larger entropies correspond to higher temperatures. At the same density and entropy, QMF generally predicts a lower temperature than RMF, consistent with its larger entropy at fixed temperature. Set 2 likewise yields lower temperatures than Set 1. The separation between the two parameter sets becomes more visible at larger entropies and is generally smaller in QMF than in RMF.

\subsection{Specific Heat ($C_V$ \& $C_P$)}

The specific heats per baryon at constant volume and constant pressure characterize the thermal response under different mechanical constraints. 
Following Ref.~\citep{2015PhRvC..92b5801C}, 
they are written as
\begin{equation}
  C_V=\frac{1}{\rho_{\rm b}}\left.\frac{\partial\varepsilon}{\partial T}\right|_{\rho_{\rm b}},
  \label{eq:CV}
\end{equation}
and 
\begin{equation}
  C_P=C_V+\frac{T}{\rho_{\rm b}^{2}}\,
  \frac{\left(\left.\partial P/\partial T\right|_{\rho_{\rm b}}\right)^{2}}
       {\left.\partial P/\partial \rho_{\rm b}\right|_{T}}.
  \label{eq:CP}
\end{equation}

\begin{figure*}
{\centering
\includegraphics[width=0.95\textwidth]{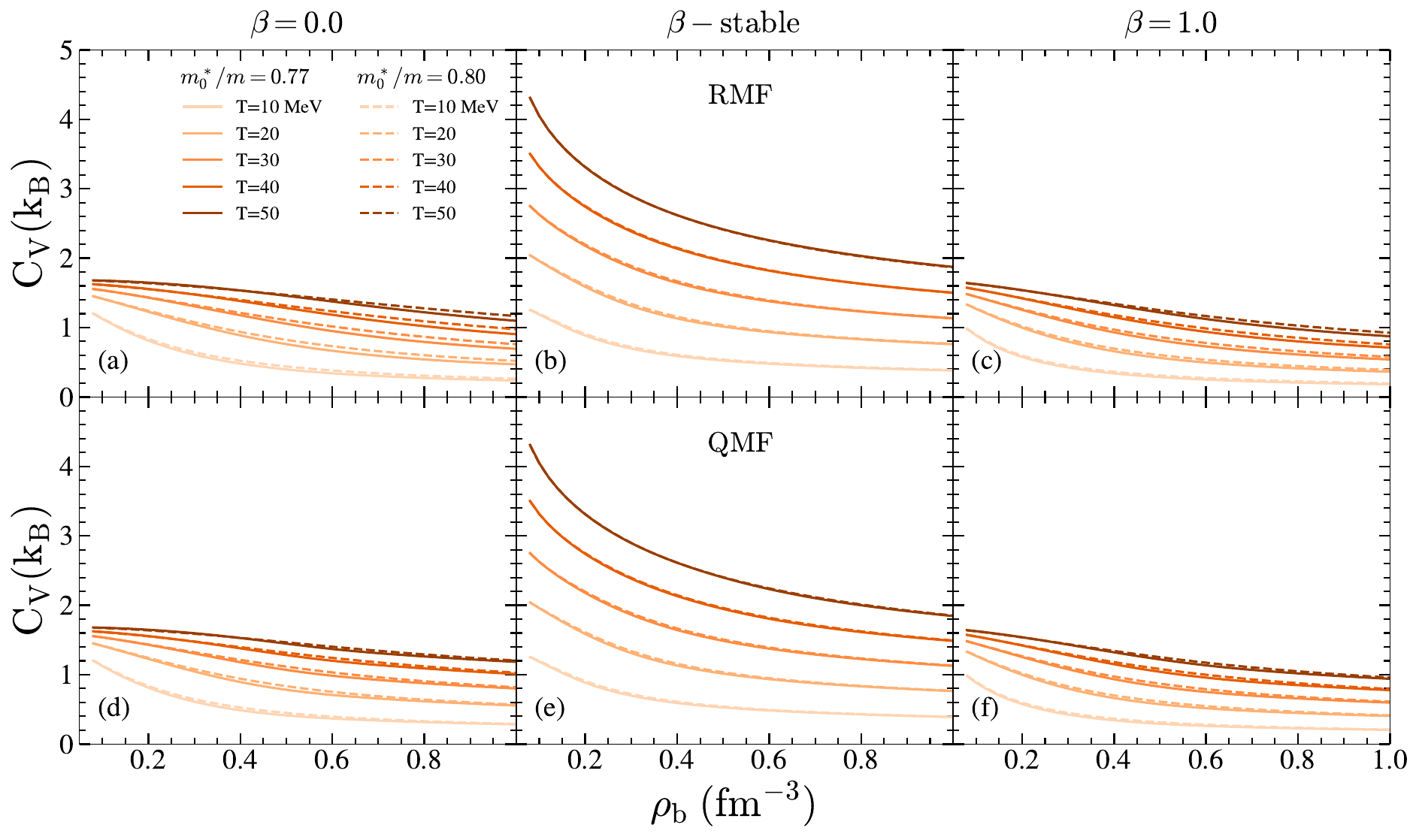}
  \caption{
   Specific heat per baryon at constant volume, $C_V$, in units of $k_{\rm B}$, as a function of baryon density $\rho_{\rm b}$ in the RMF (upper row) and QMF (lower row) models. Columns show symmetric nuclear matter, $\beta$-equilibrium matter, and pure neutron matter from left to right. Solid and dashed curves denote Sets 1 and 2, respectively; progressively darker colors indicate temperatures from 10 to 50 MeV in steps of 10 MeV.}
  \label{fig:CV}
  }\vspace{-0.1cm}
\end{figure*}

Figure \ref{fig:CV} shows $C_V$ as a function of baryon density. Both models predict larger values at higher temperatures and a broadly decreasing density dependence. In symmetric nuclear matter and pure neutron matter, the higher-temperature curves exhibit a shallow maximum at low density. By contrast, $C_V$ decreases monotonically with density for $\beta$-equilibrium matter over the range shown. The larger values in $\beta$-equilibrium matter reflect both its equilibrium composition and the additional leptonic contributions.
The QMF–RMF differences in $C_V$ are modest, with QMF generally yielding somewhat larger values at high density. The sensitivity to the effective-mass ratio at saturation is also relatively weak. 

The decrease of $C_V$ with increasing density can be understood within the Fermi-liquid framework~\citep{2001PhRvC..63e4604S, 2012PhRvC..85a4318A}.
Although the Fermi momentum increases with density, the effective mass of nucleons decreases due to the enhanced mean-field interaction at high densities.
Compared with pure neutron matter, symmetric matter has a larger $C_V$. 
In addition, higher temperatures correspond to larger $C_V$, while the influence of the effective mass $m^*_0/m$ on $C_V$ is not significant.
The increase of $C_V$ with temperature originates from the larger thermal excitation window around the Fermi surface, where more nucleons can participate in excitations.

\begin{figure*}
{\centering
\includegraphics[width=0.95\textwidth]{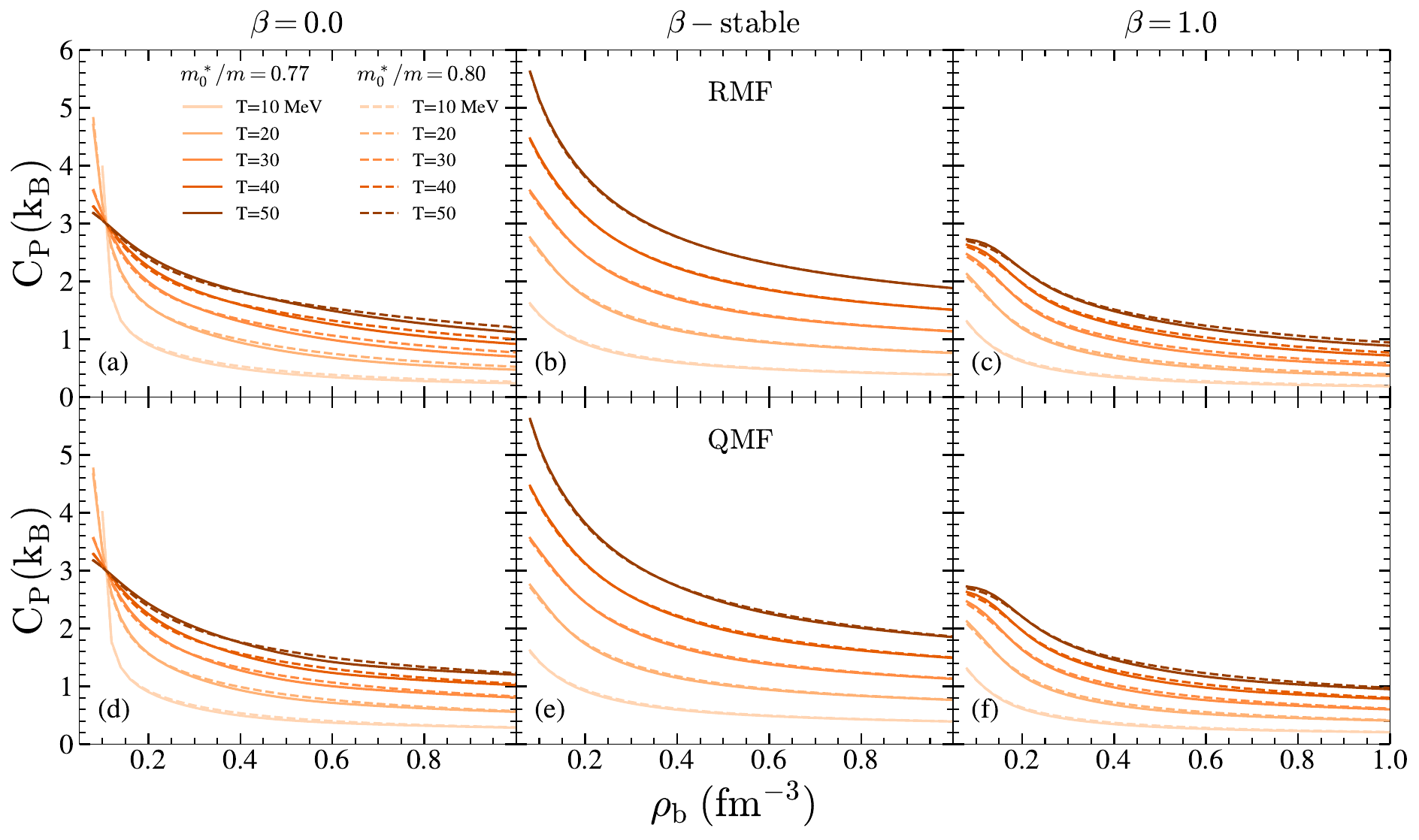}
  \caption{
  Specific heat per baryon at constant pressure, $C_P$, in units of $k_{\rm B}$, as a function of baryon density $\rho_{\rm b}$ in the RMF (upper row) and QMF (lower row) models. Columns show symmetric nuclear matter, $\beta$-equilibrium matter, and pure neutron matter from left to right. Solid and dashed curves denote Sets 1 and 2, respectively; progressively darker colors indicate temperatures from 10 to 50 MeV in steps of 10 MeV.}
  \label{fig:CP}
  }\vspace{-0.1cm}
\end{figure*}

Figure \ref{fig:CP} shows $C_P$. On mechanically stable branches, Eq. (\ref{eq:CP}) implies $C_P \ge C_V$. 
For both the RMF and QMF models, a characteristic enhancement of $C_ P$ appears in symmetric nuclear matter at low temperatures, where $C_ P$ rises sharply in the low-density region. 
This enhancement is associated with the proximity to the nuclear liquid-gas phase transition, where the EoS approaches the instability region characterized by $\partial P/\partial \rho_{\rm b}=0$~\citep{1999PhRvC..59..682B}. 
As the temperature increases, the liquid-gas phase transition becomes gradually smeared out, and the enhancement of $C_P$ is significantly reduced. 
At high densities, $C_P$ exhibits a density dependence similar to that of $C_V$, mainly governed by the properties of degenerate nuclear matter.

\begin{figure*}
{\centering
\includegraphics[width=0.95\textwidth]{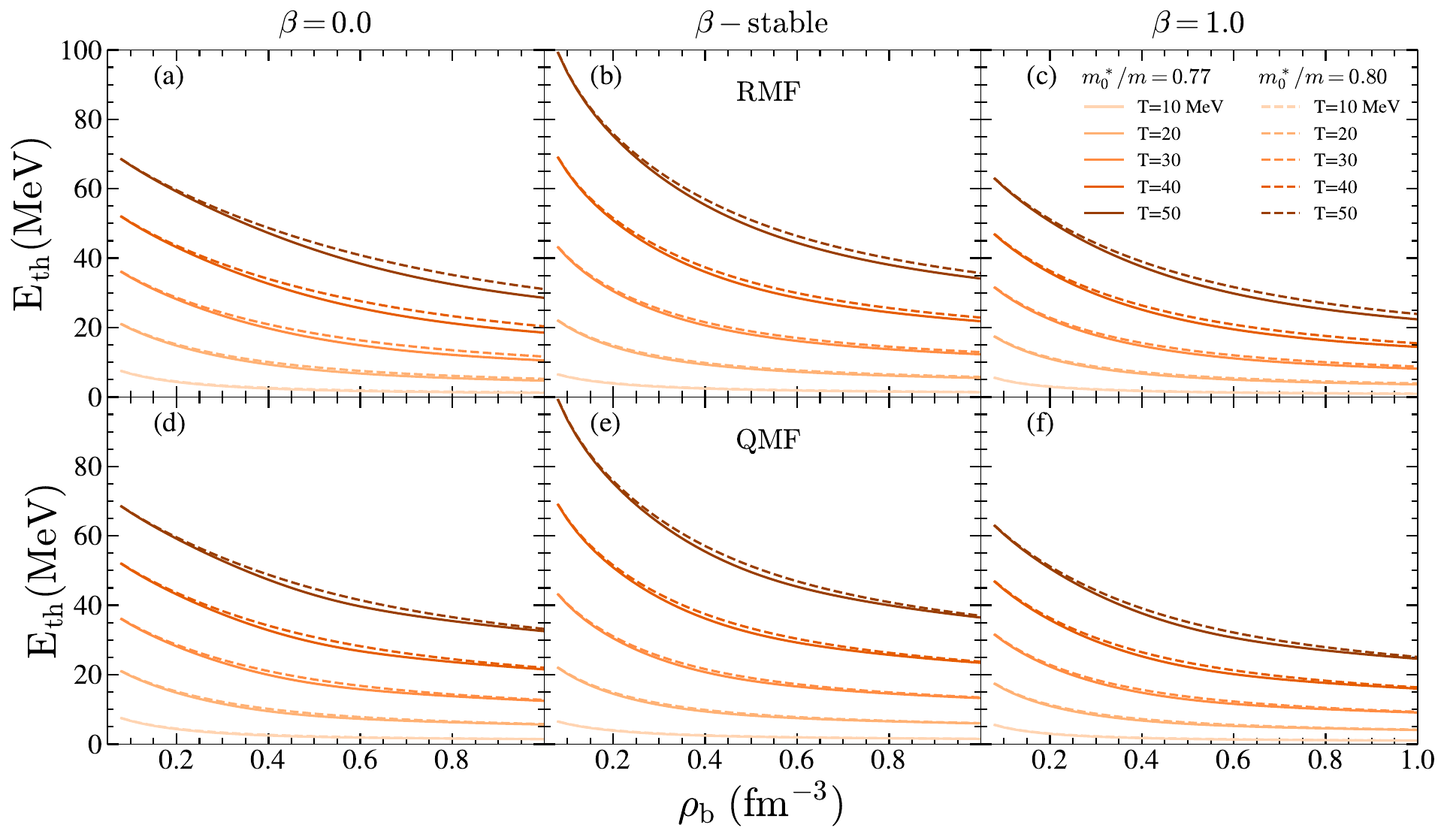}
  \caption{Thermal energy per baryon $E_{\rm{th}}$ as a function of baryon density $\rho_{\rm b}$ in the RMF (upper row) and QMF (lower row) models. The left, middle, and right columns show symmetric nuclear matter ($\beta = 0$), $\beta$-equilibrium matter, and pure neutron matter ($\beta = 1$), respectively. Solid and dashed curves denote Sets 1 and 2, corresponding to effective-mass ratios at saturation $m_0^*/m$ = 0.77 and 0.80. Progressively darker colors indicate $T$ = 10, 20, 30, 40, and 50 MeV. Leptonic contributions are included in the $\beta$-equilibrium results.}
  \label{fig:Eth}
  }\vspace{-0.1cm}
\end{figure*}

\section{Thermal Effects and Relevance to NS Mergers}

In this section, we examine the thermal contributions to the energy density and pressure, discuss the thermal index and its compact analytic parametrization, and outline the implications for BNS merger simulations.

\subsection{Thermal Energy ($E_{\rm th}$) and Pressure ($P_{\rm th}$)}

\begin{figure*}
{\centering
\includegraphics[width=0.95\textwidth]{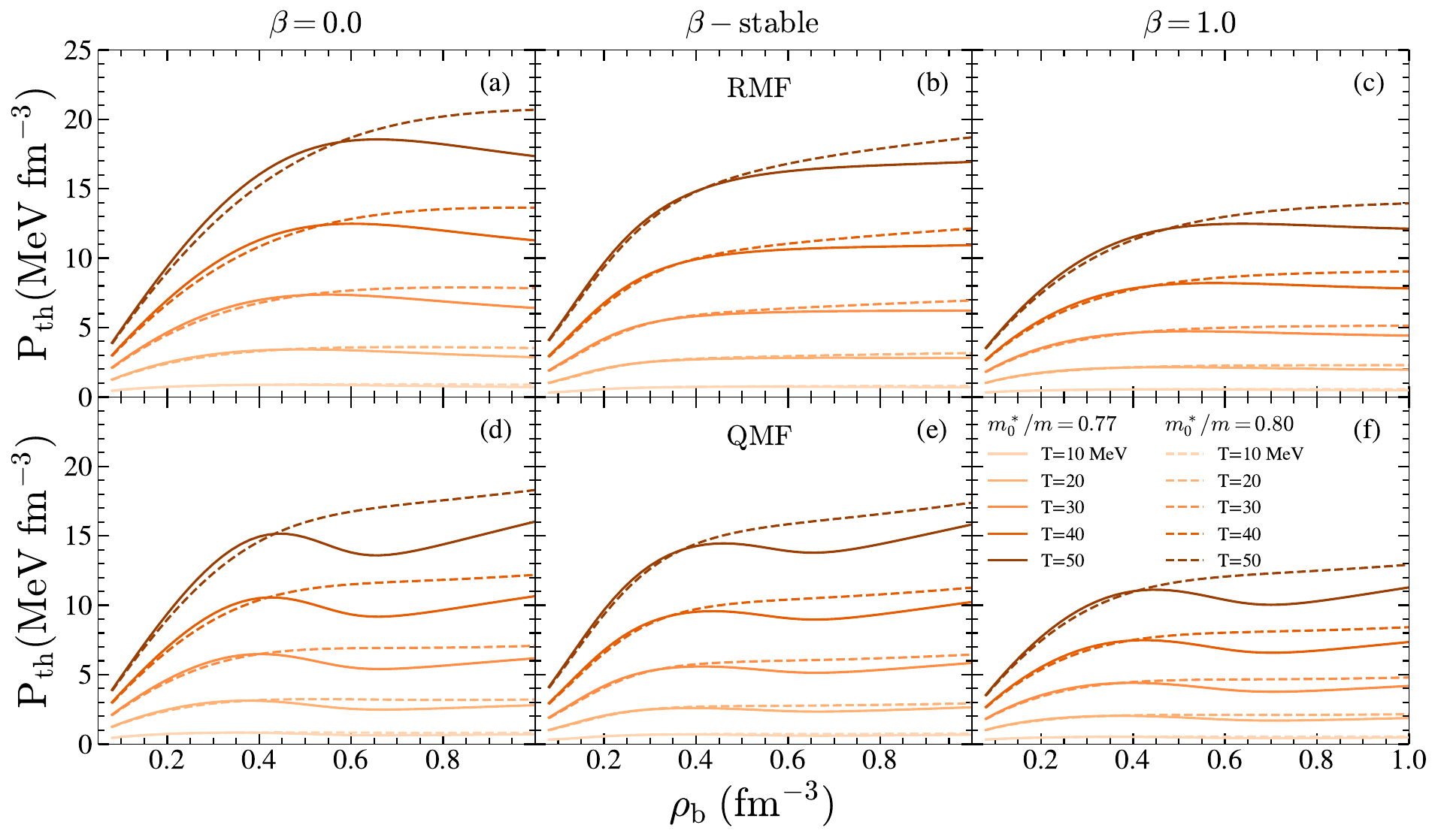}
  \caption{Thermal pressure $P_{\rm{th}}$ as a function of baryon density $\rho_{\rm b}$ in the RMF (upper row) and QMF (lower row) models. Columns show symmetric nuclear matter, $\beta$-equilibrium matter, and pure neutron matter from left to right. Solid and dashed curves denote Sets 1 and 2, respectively; progressively darker colors indicate temperatures from 10 to 50 MeV in steps of 10 MeV. Leptonic contributions are included in the $\beta$-equilibrium results.}
  \label{fig:Pth}
  }\vspace{-0.1cm}
\end{figure*}

Finite-temperature effects can be studied by isolating the thermal part of the various functions of interest, defined as the difference between the finite-$T$ and $T = 0$ expressions for a given quantity.
To quantify the effect of temperature on the study of the state equation, the thermal energy per baryon $E_{\rm th}=\varepsilon_{\rm th}/\rho_{\rm b}$ and the thermal pressure $P_{\rm th}$ from the RMF and QMF models are shown as functions of baryon density in Figs. \ref{fig:Eth} and \ref{fig:Pth}, respectively. 
The thermal energy defined as
\begin{eqnarray}
    E_{\rm th} = E(\rho_{\rm b},\beta,T) - E(\rho_{\rm b},\beta,0),
\end{eqnarray}
and the thermal pressure defined as
\begin{eqnarray}
    P_{\rm th} = P(\rho_{\rm b},\beta,T) - P(\rho_{\rm b},\beta,0).
\end{eqnarray}
Note that for matter in $\beta$ equilibrium, the quantities are evaluated at the self-consistent equilibrium composition at each temperature, with nucleonic and leptonic contributions included consistently. Thus, the equilibrium-sequence subtraction is performed at the same baryon density but generally different compositions and includes both thermal excitation and temperature-induced composition changes.

The thermal energy obtained from both models exhibits a decreasing trend with increasing density. 
Thermal effects generally enhance the thermal energy. 
Comparing symmetric nuclear matter and pure neutron matter, the influence of the isospin asymmetry is weak.
Similar to the earlier results for entropy and specific heat, the influence of the effective mass $m^*_0/m$ on the thermal energy becomes apparent only in the high-density region. 
A notable distinction arises between the RMF and QMF results. 
The thermal energy in the QMF model is slightly higher than in the RMF model, particularly in the high-temperature and high-density region.
We attribute these behaviors to the significantly different behaviors of the effective masses (both their magnitudes and density dependencies) in these two models (see Fig. \ref{fig:gsig&meff}).

The thermal pressure $P_{\rm th}$ as a function of baryon density is presented in Fig. \ref{fig:Pth} for symmetric nuclear matter ($\beta$=0), $\beta$-equilibrium matter, and pure neutron matter ($\beta$=1) within the RMF and QMF models. 
Roughly speaking, $P_{\rm th}$ increases with increasing density and temperature for all cases. 
At low densities, the thermal pressure exhibits an approximately linear dependence on density. 
With increasing density, the growth of $P_{\rm th}$ gradually slows down and may even turn into a decrease. 
For symmetric nuclear matter, $P_{\rm th}$ is systematically larger than that of pure neutron matter and $\beta$-equilibrium matter.
Comparing the RMF and QMF results, both models exhibit similar qualitative trends. 
However, at high densities, the thermal pressures calculated with different effective masses differ substantially. 
A noticeable differences appear at high densities, especially for Set 1 ($m^*_0/m=0.77$), where the QMF results show a more pronounced deviation from the monotonic increase observed for Set 2 ($m^*_0/m=0.80$). 
This difference originates from the distinct density dependence of the meson-nucleon coupling strength $g_{\sigma N}$ and the nucleon effective mass in the two models (see Fig. \ref{fig:gsig&meff}).

\subsection{Thermal Index ($\Gamma_{th}$)}

\begin{figure*}
{\centering
\includegraphics[width=0.95\textwidth]{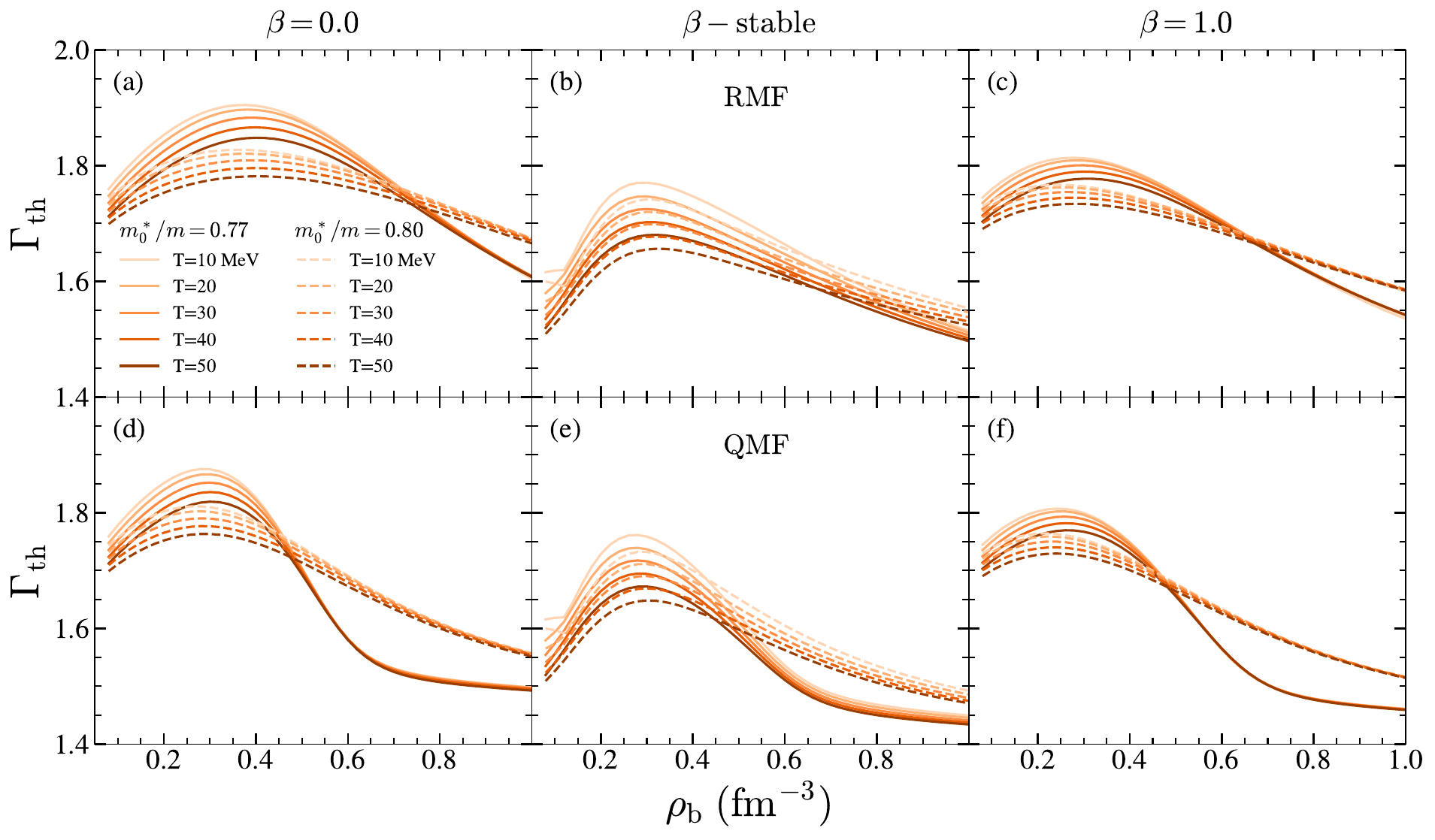}
  \caption{ Thermal index $\Gamma_{\rm{th}} = 1 + P_{\rm{th}}/\varepsilon_{\rm{th}}$ as a function of baryon density $\rho_{\rm b}$ in the RMF (upper row) and QMF (lower row) models. Columns show symmetric nuclear matter, $\beta$-equilibrium matter, and pure neutron matter from left to right. Solid and dashed curves denote Sets 1 and 2, respectively; progressively darker colors indicate temperatures from 10 to 50 MeV in steps of 10 MeV. The $\beta$-equilibrium results include nucleonic and leptonic contributions.}
  \label{fig:Gamma}
  }\vspace{-0.1cm}
\end{figure*}

\begin{table}[t]
    \centering
    \caption{Coefficients of the analytic QMF thermal-index parametrization in Eqs.~(\ref{eq:fitGamma}) and (\ref{eq:Gamfitvars}) for Sets 1 and 2. The background coefficients are $a_0=0.5680$, $a_1=0.0170$, $a_2=-0.0334$ for Set 1, and $a_0=0.5458$, $a_1=0.0163$, $a_2=-0.0290$ for Set 2. The parametrization applies to $0.08 \le \rho_{\rm b} \le 1.0~{\rm fm}^{-3}, \, 5 \le T \le 100~{\rm MeV}, \, \text{and}\,  0 \le \beta \le 1$.
    }   
    \label{tab:Gamfit}
    \begin{ruledtabular}
    \begin{tabular}{lcccc}
    & $j=1$ & $j=2$ & $j=3$ & $j=4$\\
    \hline
    \multicolumn{5}{l}{Set 1 ($m_0^{*}/m=0.77$)}\\
    $A_j$ & $0.2743$ & $-0.3782$ & $0.7371$ & $-0.0312$\\
    $w_j$ & $2.4790$ & $-3.5441$ & $-2.5325$ & $-0.4461$\\
    $p_j$ & $-0.2319$ & $0.2024$ & $0.0720$ & $0.6729$\\
    $r_j$ & $0.0665$ & $-0.3263$ & $-0.1135$ & $-2.1686$\\
    $s_j$ & $1.6313$ & $-0.8379$ & $-0.6568$ & $1.6772$\\
    \hline
    \multicolumn{5}{l}{Set 2 ($m_0^{*}/m=0.80$)}\\
    $A_j$ & $0.3586$ & $0.2108$ & $0.1960$ & $-0.0222$\\
    $w_j$ & $1.6302$ & $-1.1332$ & $-2.0385$ & $-0.6926$\\
    $p_j$ & $-0.0827$ & $-0.0589$ & $-0.0781$ & $0.9205$\\
    $r_j$ & $-0.1298$ & $0.1181$ & $0.2512$ & $-3.0106$\\
    $s_j$ & $1.2041$ & $-0.0493$ & $-1.2228$ & $2.2304$\\
    \end{tabular}
    \end{ruledtabular}
\end{table}

The inclusion of thermal effects in NS merger simulations is often treated using an effective thermal index $\Gamma_{\rm th}$ defined as 
\begin{equation}
\Gamma_{\rm th}(\rho_{\rm b}, \beta, T)
= 1 + \frac{P_{\rm th}(\rho_{\rm b}, \beta, T)}
{\varepsilon_{\rm th}(\rho_{\rm b}, \beta, T)}.
\end{equation}
It measures the thermal pressure generated per unit thermal energy density. Unlike the constant value often adopted in hybrid EoS prescriptions, the microscopic thermal index generally depends on density, temperature, and composition.

Figure \ref{fig:Gamma} shows that both models predict a broadly similar density dependence: $\Gamma_{\rm th}$ initially increases, reaches a maximum at intermediate density, and then decreases. 
It is instructive to compare our results with the analytical relations that hold in the strongly degenerate regime. 
In Landau Fermi-liquid theory, the thermal index can be written as~\citep{2015PhRvC..92b5801C}
\begin{equation}
  \Gamma_{\rm th}(\rho_{\rm b})=1+\frac{2}{3}\,
  \frac{\sum_i a_i \rho_i Q_i}{\sum_i a_i \rho_i},~
  Q_i=1-\frac{3}{2}\frac{\rho_i}{m_{L_i}^{*}}\frac{dm_{L_i}^{*}}{d\rho_{\rm b}},
  \label{eq:GamFLT}
\end{equation}
where $a_i=(\pi^2/2)m^*_{L_i}/p^2_{Fi}$ is the level-density parameter of species $i$, with $i=n,p,e,\mu$.
For pure neutron matter and for symmetric nuclear matter without leptons this reduces to
\begin{equation}
    \Gamma_{\rm th}(\rho_{\rm b})=\frac{5}{3}-\frac{\rho_{\rm b}}{m_{L_{\rm b}}^*}\frac{dm^*_{L_{\rm b}}}{d\rho_{\rm b}},\label{eq:Gam64}
\end{equation}
i.e. the thermal index is governed by the density dependence of the Landau effective mass. 
Applied to our results, Eq. (\ref{eq:Gam64}) implies that the smaller $\Gamma_{\rm th}$ obtained in the QMF model at high densities corresponds to a shallower slope, which can be verified directly against the effective masses shown in Fig. \ref{fig:gsig&meff}.
Similarly, for Set 1 ($m_0^*/m=0.77$), the effective mass varies more gradually with density in the high-density region, resulting in a lower $\Gamma_{\rm th}$ than that obtained for Set 2 ($m^*_0/m=0.80$).
The temperature dependence of $\Gamma_{\rm th}$ is relatively weak, although a moderate variation can be observed at low and intermediate densities. 
With increasing temperature, the differences among different thermal indices become more pronounced near the maximum region, but less pronounced at high densities.
A comparison among different isospin compositions shows that symmetric nuclear matter and pure neutron matter have similar qualitative behaviors, while $\beta$-equilibrium matter exhibits a slightly different density dependence due to the variation of proton fraction with density and temperature. 

\subsection{Analytic Parametrization}

For practical applications, it is desirable to obtain a compact analytical expression for $\Gamma_{\rm th}(\rho_{\rm b},\beta,T)$. 
We employ a machine-learning-assisted procedure in which a Kolmogorov--Arnold network (KAN)~\citep{2024arXiv240419756L,2025PhRvX..15d1051L} is first trained to learn the mapping $(\rho_{\rm b},T,\beta)\mapsto\Gamma_{\rm th}$. 
Previously, KANs have recently been applied to nuclear binding energies~\citep{2025PhRvC.111b4316L} and nuclear wave functions~\citep{2025PhRvC.112e4002B}, while related symbolic-regression approaches have been used to construct analytical representations of nuclear observables~\citep{2026NuPhA106823327S}. 
The learned edge functions are subsequently replaced by elementary functions from a restricted library, yielding a sparse functional representation in which only a few bounded functions survive, namely $\tanh$, $\exp(-t^2)$, and $\arctan$. 
The resulting ansatz is fitted to all $10\,560$ data points using a multi-start trust-region least-squares algorithm, with a weak regularization term used to keep the amplitudes of order unity. 
The fitted coefficients are then rounded to four decimal places, resulting in a compact analytical parametrization of the QMF thermal index. 

The resulting representation is 
\begin{eqnarray}
    \Gamma_{\rm th}(\rho_{\rm b},\beta,T)&=&1+a_0+a_1q+a_2q^2\nonumber\\
    &+&A_1\tanh(w_1\tilde x+p_1\tilde \tau+r_1q+s_1)\nonumber\\
    &+&A_2\tanh(w_2\tilde x+p_2\tilde \tau+r_2q+s_2)\nonumber\\
    &+&A_3\tanh(w_3\tilde x+p_3\tilde \tau+r_3q+s_3)\nonumber\\
    &+&A_4\tanh(w_4\tilde x+p_4\tilde \tau+r_4q+s_4)\nonumber\\
    \label{eq:fitGamma}
\end{eqnarray}
with the dimensionless variables
\begin{equation}
    \tilde x=\frac{\rho_{\rm b}/\rho_0-3.3000}{3.0000},~
    \tilde \tau=\frac{T/10-5.5000}{5.5000},~
    q=(1-\beta)^{1/3}.
    \label{eq:Gamfitvars}
\end{equation}
The coefficients are listed in Table \ref{tab:Gamfit} for the two parameter sets.
Over the ranges $0.08\le\rho_{\rm b}\le1.0~\rm fm^{-3}$, $5\le T\le 100 ~\rm MeV$ and $0\le \beta\le 1$, Eq. (\ref{eq:fitGamma}) reproduces the numerical QMF thermal index $\Gamma_{\rm th}$ with root-mean-square deviations of $3.57\times10^{-3}$ and $2.27\times10^{-3}$ for Set~1 and Set~2, respectively. The corresponding maximum relative deviations are approximately $1.13\%$ and $1.02\%$.
These expressions are well within the accuracy required by merger simulations and can be implemented directly in numerical-relativity codes. 

\subsection{Implications for NS Mergers}

Shock heating during a BNS merger can raise the matter temperature to several tens of MeV, while the remnant reaches densities above nuclear saturation~\citep{2022MNRAS.516.4792R}. 
Thermal pressure then provides additional support and can influence the remnant structure, stability, and post-merger GW emission. Previous simulations have demonstrated that these effects depend on the treatment of finite-temperature matter~\citep{2020PhRvD.102d3006F,2021PhRvD.103h3012F, 2023ApJ...952L..36F}. 

In the present work, we have extended the QMF model to finite temperature and systematically investigated its thermal properties, including the specific heats $C_{\rm V}$ and $C_{\rm P}$ and the thermal pressure. 
The results demonstrate that the thermal response of QMF matter depends on both density and temperature and differs from that of the RMF model used for comparison. 
In particular, the relatively lower thermal pressure obtained in the QMF model compared with RMF may imply weaker thermal support for a hot post-merger remnant~\citep{2023ApJ...952L..36F}. 
For otherwise similar cold-matter properties, this could lead to a more compact post-merger remnant and may affect the onset of gravitational collapse.
However, the quantitative impact of the thermal prescription on the prompt-collapse threshold depends on the full merger dynamics and requires dedicated relativistic simulations~\citep{2026PhRvD.113d3060R}.
The density and temperature dependence of $\Gamma_{\rm th}$ may also lead to different thermal pressure profiles during the evolution of the remnant, which can affect its compactness and oscillation modes.

These differences may further leave an imprint on the post-merger GW signal. 
In particular, the characteristic peak frequency $f_{\rm peak}$ is sensitive to the structure and compactness of the post-merger remnant and can therefore be affected by the stiffness of the EoS~\citep{2021PhRvD.104f3016R, 2023PhRvD.108h3029R}. 
The distinct density- and temperature-dependent thermal responses predicted by different models, as illustrated here by RMF and QMF, may therefore lead to different values of $f_{\rm peak}$. 
Moreover, The model-dependent equilibrium lepton fractions motivate further study of weak-interaction rates and composition evolution. The difference of $Y_e$ and $Y_\mu$ between two models leads to different out-of-equilibrium behaviors of the EoS, resulting in different neutrino processes and ejecta properties.
A quantitative determination of these effects requires future relativistic BNS merger simulations using fully temperature- and composition-dependent EoSs, rather than a constant-$\Gamma_{\rm th}$ approximation.
Such an investigation is currently underway and will be presented in a separate work.

\section{Conclusion}

In this work, we have extended the QMF model to finite temperature and compared its predictions with those of RMF models calibrated to the same selected nuclear-matter saturation properties. Two effective-mass ratios at saturation, 0.77 and 0.80, were considered. 
The calculations cover symmetric nuclear matter, pure neutron matter, and $\beta$-equilibrium matter.

The models exhibit similar qualitative thermodynamic trends, but their predictions increasingly differ at high density. The principal distinction is the density evolution of the effective nucleon mass: the QMF Dirac mass varies more slowly at high density than its RMF counterpart for the parameter sets studied. This difference is reflected in the entropy, thermal energy, thermal pressure, and thermal index. In particular, the degenerate-limit relation involving the density derivative of the Landau effective mass provides a useful interpretation of the thermal-index behavior.

We have also briefly discussed the astrophysical implications of the finite-temperature QMF EoS for BNS merger simulations. 
The different thermal pressures predicted by QMF and RMF may lead to different thermal support of the hot post-merger remnant and consequently affect its stability and evolution. 
The density- and temperature-dependent thermal index obtained in the present work may also provide a more microscopic alternative to the commonly adopted constant-$\Gamma_{\rm th}$ prescription in future merger simulations. 
To facilitate such applications, we provide compact analytic representations of the QMF thermal index over baryon densities of $0.08$-$1.0~ \rm fm^{-3}$, temperatures of $5$-$100$ MeV, and isospin asymmetries of 0-1. Within this domain, the reported maximum deviations are below 2\%.
A direct implementation of the finite-temperature QMF EoS in dynamical BNS merger simulations will be an important direction for future work.  

\medskip
\acknowledgments
We are thankful to the XMU neutron star group for helpful discussions.
The work is supported by the National Natural Science Foundation of China (grant Nos. 12494572, 12273028).

\bibliographystyle{unsrt}
\bibliography{ref}

@ARTICLE{2011PrPNP..66..519N,
       author = {{Nik{\v{s}}i{\'c}}, T. and {Vretenar}, D. and {Ring}, P.},
        title = "{Relativistic nuclear energy density functionals: Mean-field and beyond}",
      journal = {Progress in Particle and Nuclear Physics},
         year = 2011,
        month = jul,
       volume = {66},
       number = {3},
        pages = {519-548},
          doi = {10.1016/j.ppnp.2011.01.055},
archivePrefix = {arXiv},
       eprint = {1102.4193},
 primaryClass = {nucl-th},
       adsurl = {https://ui.adsabs.harvard.edu/abs/2011PrPNP..66..519N}
}

@ARTICLE{2000PhRvC..61d5205S,
       author = {{Shen}, H. and {Toki}, H.},
        title = "{Quark mean field model for nuclear matter and finite nuclei}",
      journal = {\prc},
         year = 2000,
        month = apr,
       volume = {61},
       number = {4},
          eid = {045205},
        pages = {045205},
          doi = {10.1103/PhysRevC.61.045205},
archivePrefix = {arXiv},
       eprint = {nucl-th/9911046},
 primaryClass = {nucl-th},
       adsurl = {https://ui.adsabs.harvard.edu/abs/2000PhRvC..61d5205S}
}

@ARTICLE{1998PhRvC..58.3749T,
       author = {{Toki}, H. and {Meyer}, U. and {Faessler}, A. and {Brockmann}, R.},
        title = "{Quark mean field model for nucleons in nuclei}",
      journal = {\prc},
         year = 1998,
        month = dec,
       volume = {58},
       number = {6},
        pages = {3749-3752},
          doi = {10.1103/PhysRevC.58.3749},
       adsurl = {https://ui.adsabs.harvard.edu/abs/1998PhRvC..58.3749T}
}

@ARTICLE{1988PhLB..200..235G,
       author = {{Guichon}, P.~A.~M.},
        title = "{A possible quark mechanism for the saturation of nuclear matter}",
      journal = {Physics Letters B},
         year = 1988,
        month = jan,
       volume = {200},
       number = {3},
        pages = {235-240},
          doi = {10.1016/0370-2693(88)90762-9},
       adsurl = {https://ui.adsabs.harvard.edu/abs/1988PhLB..200..235G}
}

@ARTICLE{2018PrPNP.100..262G,
       author = {{Guichon}, P.~A.~M. and {Stone}, J.~R. and {Thomas}, A.~W.},
        title = "{Quark-Meson-Coupling (QMC) model for finite nuclei, nuclear matter and beyond}",
      journal = {Progress in Particle and Nuclear Physics},
         year = 2018,
        month = may,
       volume = {100},
        pages = {262-297},
          doi = {10.1016/j.ppnp.2018.01.008},
archivePrefix = {arXiv},
       eprint = {1802.08368},
 primaryClass = {nucl-th},
       adsurl = {https://ui.adsabs.harvard.edu/abs/2018PrPNP.100..262G}
}

@ARTICLE{2022NatRP...4..237Y,
       author = {{Yunes}, Nicol{\'a}s and {Miller}, M. Coleman and {Yagi}, Kent},
        title = "{Gravitational-wave and X-ray probes of the neutron star equation of state}",
      journal = {Nature Reviews Physics},
         year = 2022,
        month = feb,
       volume = {4},
       number = {4},
        pages = {237-246},
          doi = {10.1038/s42254-022-00420-y},
archivePrefix = {arXiv},
       eprint = {2202.04117},
 primaryClass = {gr-qc},
       adsurl = {https://ui.adsabs.harvard.edu/abs/2022NatRP...4..237Y}
}

@ARTICLE{2021ApJ...913...27L,
       author = {{Li}, Ang and {Miao}, Zhiqiang and {Han}, Sophia and {Zhang}, Bing},
        title = "{Constraints on the Maximum Mass of Neutron Stars with a Quark Core from GW170817 and NICER PSR J0030+0451 Data}",
      journal = {\apj},
         year = 2021,
        month = may,
       volume = {913},
       number = {1},
          eid = {27},
        pages = {27},
          doi = {10.3847/1538-4357/abf355},
archivePrefix = {arXiv},
       eprint = {2103.15119},
 primaryClass = {astro-ph.HE},
       adsurl = {https://ui.adsabs.harvard.edu/abs/2021ApJ...913...27L}
}

@ARTICLE{2019Univ....5..159L,
       author = {{Lattimer}, James M.},
        title = "{Neutron Star Mass and Radius Measurements}",
      journal = {Universe},
         year = 2019,
        month = jun,
       volume = {5},
       number = {7},
          eid = {159},
        pages = {159},
          doi = {10.3390/universe5070159},
       adsurl = {https://ui.adsabs.harvard.edu/abs/2019Univ....5..159L}
}

@ARTICLE{2023PhRvD.108f3032B,
       author = {{Blacker}, Sebastian and {Bauswein}, Andreas and {Typel}, Stefan},
        title = "{Exploring thermal effects of the hadron-quark matter transition in neutron star mergers}",
      journal = {\prd},
         year = 2023,
        month = sep,
       volume = {108},
       number = {6},
          eid = {063032},
        pages = {063032},
          doi = {10.1103/PhysRevD.108.063032},
archivePrefix = {arXiv},
       eprint = {2304.01971},
 primaryClass = {astro-ph.HE},
       adsurl = {https://ui.adsabs.harvard.edu/abs/2023PhRvD.108f3032B}
}

@ARTICLE{2023ApJ...943..163Z,
       author = {{Zhu}, Zhenyu and {Li}, Ang and {Liu}, Tong},
        title = "{A Bayesian Inference of a Relativistic Mean-field Model of Neutron Star Matter from Observations of NICER and GW170817/AT2017gfo}",
      journal = {\apj},
         year = 2023,
        month = feb,
       volume = {943},
       number = {2},
          eid = {163},
        pages = {163},
          doi = {10.3847/1538-4357/acac1f},
archivePrefix = {arXiv},
       eprint = {2211.02007},
 primaryClass = {astro-ph.HE},
       adsurl = {https://ui.adsabs.harvard.edu/abs/2023ApJ...943..163Z}
}

@ARTICLE{2018PhRvC..97c5805Z,
       author = {{Zhu}, Zhen-Yu and {Li}, Ang},
        title = "{Effects of the nucleon radius on neutron stars in a quark mean field model}",
      journal = {\prc},
         year = 2018,
        month = mar,
       volume = {97},
       number = {3},
          eid = {035805},
        pages = {035805},
          doi = {10.1103/PhysRevC.97.035805},
archivePrefix = {arXiv},
       eprint = {1802.07441},
 primaryClass = {nucl-th},
       adsurl = {https://ui.adsabs.harvard.edu/abs/2018PhRvC..97c5805Z}
}

@article{BOGUTA1977413,
title = {Relativistic calculation of nuclear matter and the nuclear surface},
journal = {Nuclear Physics A},
volume = {292},
number = {3},
pages = {413-428},
year = {1977},
issn = {0375-9474},
doi = {https://doi.org/10.1016/0375-9474(77)90626-1},
url = {https://www.sciencedirect.com/science/article/pii/0375947477906261},
author = {J. Boguta and A.R. Bodmer}
}

@ARTICLE{2001PhRvL..86.5647H,
       author = {{Horowitz}, C.~J. and {Piekarewicz}, J.},
        title = "{Neutron Star Structure and the Neutron Radius of $^{208}$Pb}",
      journal = {\prl},
         year = 2001,
        month = jun,
       volume = {86},
       number = {25},
        pages = {5647-5650},
          doi = {10.1103/PhysRevLett.86.5647},
archivePrefix = {arXiv},
       eprint = {astro-ph/0010227},
 primaryClass = {astro-ph},
       adsurl = {https://ui.adsabs.harvard.edu/abs/2001PhRvL..86.5647H}
}

@ARTICLE{2020PhRvD.102d3006F,
       author = {{Figura}, A. and {Lu}, Jia-Jing and {Burgio}, G.~F. and {Li}, Zeng-Hua and {Schulze}, H.-J.},
        title = "{Hybrid equation of state approach in binary neutron-star merger simulations}",
      journal = {\prd},
         year = 2020,
        month = aug,
       volume = {102},
       number = {4},
          eid = {043006},
        pages = {043006},
          doi = {10.1103/PhysRevD.102.043006},
archivePrefix = {arXiv},
       eprint = {2005.08691},
 primaryClass = {gr-qc},
       adsurl = {https://ui.adsabs.harvard.edu/abs/2020PhRvD.102d3006F}
}

@ARTICLE{2023PhRvD.108h3029R,
       author = {{Raithel}, Carolyn A. and {Paschalidis}, Vasileios},
        title = "{Influence of stellar compactness on finite-temperature effects in neutron star merger simulations}",
      journal = {\prd},
         year = 2023,
        month = oct,
       volume = {108},
       number = {8},
          eid = {083029},
        pages = {083029},
          doi = {10.1103/PhysRevD.108.083029},
archivePrefix = {arXiv},
       eprint = {2306.13144},
 primaryClass = {astro-ph.HE},
       adsurl = {https://ui.adsabs.harvard.edu/abs/2023PhRvD.108h3029R}
}

@ARTICLE{2015PhRvC..92b5801C,
       author = {{Constantinou}, Constantinos and {Muccioli}, Brian and {Prakash}, Madappa and {Lattimer}, James M.},
        title = "{Thermal properties of hot and dense matter with finite range interactions}",
      journal = {\prc},
         year = 2015,
        month = aug,
       volume = {92},
       number = {2},
          eid = {025801},
        pages = {025801},
          doi = {10.1103/PhysRevC.92.025801},
archivePrefix = {arXiv},
       eprint = {1504.03982},
 primaryClass = {astro-ph.SR},
       adsurl = {https://ui.adsabs.harvard.edu/abs/2015PhRvC..92b5801C}
}

@ARTICLE{2021PhRvD.103h3012F,
       author = {{Figura}, A. and {Li}, Fan and {Lu}, Jia-Jing and {Burgio}, G.~F. and {Li}, Zeng-Hua and {Schulze}, H.-J.},
        title = "{Binary neutron star merger simulations with hot microscopic equations of state}",
      journal = {\prd},
         year = 2021,
        month = apr,
       volume = {103},
       number = {8},
          eid = {083012},
        pages = {083012},
          doi = {10.1103/PhysRevD.103.083012},
archivePrefix = {arXiv},
       eprint = {2103.02365},
 primaryClass = {gr-qc},
       adsurl = {https://ui.adsabs.harvard.edu/abs/2021PhRvD.103h3012F}
}

@ARTICLE{2017PhRvL.119p1101A,
       author = {{Abbott}, B.~P. and {Abbott}, R. and {Abbott}, T.~D. and {Acernese}, F. and {Ackley}, K. and {Adams}, C. and {Adams}, T. and {Addesso}, P. and {Adhikari}, R.~X. and {Adya}, V.~B. and {Affeldt}, C. and {Afrough}, M. and {Agarwal}, B. and {Agathos}, M. and {Agatsuma}, K. and {Aggarwal}, N. and {Aguiar}, O.~D. and {Aiello}, L. and {Ain}, A. and {Ajith}, P. and {Allen}, B. and {Allen}, G. and {Allocca}, A. and {Altin}, P.~A. and {Amato}, A. and {Ananyeva}, A. and {Anderson}, S.~B. and {Anderson}, W.~G. and {Angelova}, S.~V. and {Antier}, S. and {Appert}, S. and {Arai}, K. and {Araya}, M.~C. and {Areeda}, J.~S. and {Arnaud}, N. and {Arun}, K.~G. and {Ascenzi}, S. and {Ashton}, G. and {Ast}, M. and {Aston}, S.~M. and {Astone}, P. and {Atallah}, D.~V. and {Aufmuth}, P. and {Aulbert}, C. and {AultONeal}, K. and {Austin}, C. and {Avila-Alvarez}, A. and {Babak}, S. and {Bacon}, P. and {Bader}, M.~K.~M. and {Bae}, S. and {Bailes}, M. and {Baker}, P.~T. and {Baldaccini}, F. and {Ballardin}, G. and {Ballmer}, S.~W. and {Banagiri}, S. and {Barayoga}, J.~C. and {Barclay}, S.~E. and {Barish}, B.~C. and {Barker}, D. and {Barkett}, K. and {Barone}, F. and {Barr}, B. and {Barsotti}, L. and {Barsuglia}, M. and {Barta}, D. and {Barthelmy}, S.~D. and {Bartlett}, J. and {Bartos}, I. and {Bassiri}, R. and {Basti}, A. and {Batch}, J.~C. and {Bawaj}, M. and {Bayley}, J.~C. and {Bazzan}, M. and {B{\'e}csy}, B. and {Beer}, C. and {Bejger}, M. and {Belahcene}, I. and {Bell}, A.~S. and {Berger}, B.~K. and {Bergmann}, G. and {Bernuzzi}, S. and {Bero}, J.~J. and {Berry}, C.~P.~L. and {Bersanetti}, D. and {Bertolini}, A. and {Betzwieser}, J. and {Bhagwat}, S. and {Bhandare}, R. and {Bilenko}, I.~A. and {Billingsley}, G. and {Billman}, C.~R. and {Birch}, J. and {Birney}, R. and {Birnholtz}, O. and {Biscans}, S. and {Biscoveanu}, S. and {Bisht}, A. and {Bitossi}, M. and {Biwer}, C. and {Bizouard}, M.~A. and {Blackburn}, J.~K. and {Blackman}, J. and {Blair}, C.~D. and {Blair}, D.~G. and {Blair}, R.~M. and {Bloemen}, S. and {Bock}, O. and {Bode}, N. and {Boer}, M. and {Bogaert}, G. and {Bohe}, A. and {Bondu}, F. and {Bonilla}, E. and {Bonnand}, R. and {Boom}, B.~A. and {Bork}, R. and {Boschi}, V. and {Bose}, S. and {Bossie}, K. and {Bouffanais}, Y. and {Bozzi}, A. and {Bradaschia}, C. and {Brady}, P.~R. and {Branchesi}, M. and {Brau}, J.~E. and {Briant}, T. and {Brillet}, A. and {Brinkmann}, M. and {Brisson}, V. and {Brockill}, P. and {Broida}, J.~E. and {Brooks}, A.~F. and {Brown}, D.~A. and {Brown}, D.~D. and {Brunett}, S. and {Buchanan}, C.~C. and {Buikema}, A. and {Bulik}, T. and {Bulten}, H.~J. and {Buonanno}, A. and {Buskulic}, D. and {Buy}, C. and {Byer}, R.~L. and {Cabero}, M. and {Cadonati}, L. and {Cagnoli}, G. and {Cahillane}, C. and {Calder{\'o}n Bustillo}, J. and {Callister}, T.~A. and {Calloni}, E. and {Camp}, J.~B. and {Canepa}, M. and {Canizares}, P. and {Cannon}, K.~C. and {Cao}, H. and {Cao}, J. and {Capano}, C.~D. and {Capocasa}, E. and {Carbognani}, F. and {Caride}, S. and {Carney}, M.~F. and {Carullo}, G. and {Casanueva Diaz}, J. and {Casentini}, C. and {Caudill}, S. and {Cavagli{\`a}}, M. and {Cavalier}, F. and {Cavalieri}, R. and {Cella}, G. and {Cepeda}, C.~B. and {Cerd{\'a}-Dur{\'a}n}, P. and {Cerretani}, G. and {Cesarini}, E. and {Chamberlin}, S.~J. and {Chan}, M. and {Chao}, S. and {Charlton}, P. and {Chase}, E. and {Chassande-Mottin}, E. and {Chatterjee}, D. and {Chatziioannou}, K. and {Cheeseboro}, B.~D. and {Chen}, H.~Y. and {Chen}, X. and {Chen}, Y. and {Cheng}, H.-P. and {Chia}, H. and {Chincarini}, A. and {Chiummo}, A. and {Chmiel}, T. and {Cho}, H.~S. and {Cho}, M. and {Chow}, J.~H. and {Christensen}, N. and {Chu}, Q. and {Chua}, A.~J.~K. and {Chua}, S.},
        title = "{GW170817: Observation of Gravitational Waves from a Binary Neutron Star Inspiral}",
      journal = {\prl},
         year = 2017,
        month = oct,
       volume = {119},
       number = {16},
          eid = {161101},
        pages = {161101},
          doi = {10.1103/PhysRevLett.119.161101},
archivePrefix = {arXiv},
       eprint = {1710.05832},
 primaryClass = {gr-qc},
       adsurl = {https://ui.adsabs.harvard.edu/abs/2017PhRvL.119p1101A}
}

@ARTICLE{2018PhRvL.121p1101A,
       author = {{Abbott}, B.~P. and {Abbott}, R. and {Abbott}, T.~D. and {Acernese}, F. and {Ackley}, K. and {Adams}, C. and {Adams}, T. and {Addesso}, P. and {Adhikari}, R.~X. and {Adya}, V.~B. and {Affeldt}, C. and {Agarwal}, B. and {Agathos}, M. and {Agatsuma}, K. and {Aggarwal}, N. and {Aguiar}, O.~D. and {Aiello}, L. and {Ain}, A. and {Ajith}, P. and {Allen}, B. and {Allen}, G. and {Allocca}, A. and {Aloy}, M.~A. and {Altin}, P.~A. and {Amato}, A. and {Ananyeva}, A. and {Anderson}, S.~B. and {Anderson}, W.~G. and {Angelova}, S.~V. and {Antier}, S. and {Appert}, S. and {Arai}, K. and {Araya}, M.~C. and {Areeda}, J.~S. and {Ar{\`e}ne}, M. and {Arnaud}, N. and {Arun}, K.~G. and {Ascenzi}, S. and {Ashton}, G. and {Ast}, M. and {Aston}, S.~M. and {Astone}, P. and {Atallah}, D.~V. and {Aubin}, F. and {Aufmuth}, P. and {Aulbert}, C. and {AultONeal}, K. and {Austin}, C. and {Avila-Alvarez}, A. and {Babak}, S. and {Bacon}, P. and {Badaracco}, F. and {Bader}, M.~K.~M. and {Bae}, S. and {Baker}, P.~T. and {Baldaccini}, F. and {Ballardin}, G. and {Ballmer}, S.~W. and {Banagiri}, S. and {Barayoga}, J.~C. and {Barclay}, S.~E. and {Barish}, B.~C. and {Barker}, D. and {Barkett}, K. and {Barnum}, S. and {Barone}, F. and {Barr}, B. and {Barsotti}, L. and {Barsuglia}, M. and {Barta}, D. and {Bartlett}, J. and {Bartos}, I. and {Bassiri}, R. and {Basti}, A. and {Batch}, J.~C. and {Bawaj}, M. and {Bayley}, J.~C. and {Bazzan}, M. and {B{\'e}csy}, B. and {Beer}, C. and {Bejger}, M. and {Belahcene}, I. and {Bell}, A.~S. and {Beniwal}, D. and {Bensch}, M. and {Berger}, B.~K. and {Bergmann}, G. and {Bernuzzi}, S. and {Bero}, J.~J. and {Berry}, C.~P.~L. and {Bersanetti}, D. and {Bertolini}, A. and {Betzwieser}, J. and {Bhandare}, R. and {Bilenko}, I.~A. and {Bilgili}, S.~A. and {Billingsley}, G. and {Billman}, C.~R. and {Birch}, J. and {Birney}, R. and {Birnholtz}, O. and {Biscans}, S. and {Biscoveanu}, S. and {Bisht}, A. and {Bitossi}, M. and {Bizouard}, M.~A. and {Blackburn}, J.~K. and {Blackman}, J. and {Blair}, C.~D. and {Blair}, D.~G. and {Blair}, R.~M. and {Bloemen}, S. and {Bock}, O. and {Bode}, N. and {Boer}, M. and {Boetzel}, Y. and {Bogaert}, G. and {Bohe}, A. and {Bondu}, F. and {Bonilla}, E. and {Bonnand}, R. and {Booker}, P. and {Boom}, B.~A. and {Booth}, C.~D. and {Bork}, R. and {Boschi}, V. and {Bose}, S. and {Bossie}, K. and {Bossilkov}, V. and {Bosveld}, J. and {Bouffanais}, Y. and {Bozzi}, A. and {Bradaschia}, C. and {Brady}, P.~R. and {Bramley}, A. and {Branchesi}, M. and {Brau}, J.~E. and {Briant}, T. and {Brighenti}, F. and {Brillet}, A. and {Brinkmann}, M. and {Brisson}, V. and {Brockill}, P. and {Brooks}, A.~F. and {Brown}, D.~D. and {Brunett}, S. and {Buchanan}, C.~C. and {Buikema}, A. and {Bulik}, T. and {Bulten}, H.~J. and {Buonanno}, A. and {Buskulic}, D. and {Buy}, C. and {Byer}, R.~L. and {Cabero}, M. and {Cadonati}, L. and {Cagnoli}, G. and {Cahillane}, C. and {Calder{\'o}n Bustillo}, J. and {Callister}, T.~A. and {Calloni}, E. and {Camp}, J.~B. and {Canepa}, M. and {Canizares}, P. and {Cannon}, K.~C. and {Cao}, H. and {Cao}, J. and {Capano}, C.~D. and {Capocasa}, E. and {Carbognani}, F. and {Caride}, S. and {Carney}, M.~F. and {Carullo}, G. and {Casanueva Diaz}, J. and {Casentini}, C. and {Caudill}, S. and {Cavagli{\`a}}, M. and {Cavalier}, F. and {Cavalieri}, R. and {Cella}, G. and {Cepeda}, C.~B. and {Cerd{\'a}-Dur{\'a}n}, P. and {Cerretani}, G. and {Cesarini}, E. and {Chaibi}, O. and {Chamberlin}, S.~J. and {Chan}, M. and {Chao}, S. and {Charlton}, P. and {Chase}, E. and {Chassande-Mottin}, E. and {Chatterjee}, D. and {Chatziioannou}, K. and {Cheeseboro}, B.~D. and {Chen}, H.~Y. and {Chen}, X. and {Chen}, Y. and {Cheng}, H.-P. and {Chia}, H.~Y. and {Chincarini}, A.},
        title = "{GW170817: Measurements of Neutron Star Radii and Equation of State}",
      journal = {\prl},
         year = 2018,
        month = oct,
       volume = {121},
       number = {16},
          eid = {161101},
        pages = {161101},
          doi = {10.1103/PhysRevLett.121.161101},
archivePrefix = {arXiv},
       eprint = {1805.11581},
 primaryClass = {gr-qc},
       adsurl = {https://ui.adsabs.harvard.edu/abs/2018PhRvL.121p1101A}
}

@ARTICLE{2019ApJ...887L..21R,
       author = {{Riley}, T.~E. and {Watts}, A.~L. and {Bogdanov}, S. and {Ray}, P.~S. and {Ludlam}, R.~M. and {Guillot}, S. and {Arzoumanian}, Z. and {Baker}, C.~L. and {Bilous}, A.~V. and {Chakrabarty}, D. and {Gendreau}, K.~C. and {Harding}, A.~K. and {Ho}, W.~C.~G. and {Lattimer}, J.~M. and {Morsink}, S.~M. and {Strohmayer}, T.~E.},
        title = "{A NICER View of PSR J0030+0451: Millisecond Pulsar Parameter Estimation}",
      journal = {\apjl},
         year = 2019,
        month = dec,
       volume = {887},
       number = {1},
          eid = {L21},
        pages = {L21},
          doi = {10.3847/2041-8213/ab481c},
archivePrefix = {arXiv},
       eprint = {1912.05702},
 primaryClass = {astro-ph.HE},
       adsurl = {https://ui.adsabs.harvard.edu/abs/2019ApJ...887L..21R}
}

@ARTICLE{2019ApJ...887L..24M,
       author = {{Miller}, M.~C. and {Lamb}, F.~K. and {Dittmann}, A.~J. and {Bogdanov}, S. and {Arzoumanian}, Z. and {Gendreau}, K.~C. and {Guillot}, S. and {Harding}, A.~K. and {Ho}, W.~C.~G. and {Lattimer}, J.~M. and {Ludlam}, R.~M. and {Mahmoodifar}, S. and {Morsink}, S.~M. and {Ray}, P.~S. and {Strohmayer}, T.~E. and {Wood}, K.~S. and {Enoto}, T. and {Foster}, R. and {Okajima}, T. and {Prigozhin}, G. and {Soong}, Y.},
        title = "{PSR J0030+0451 Mass and Radius from NICER Data and Implications for the Properties of Neutron Star Matter}",
      journal = {\apjl},
         year = 2019,
        month = dec,
       volume = {887},
       number = {1},
          eid = {L24},
        pages = {L24},
          doi = {10.3847/2041-8213/ab50c5},
archivePrefix = {arXiv},
       eprint = {1912.05705},
 primaryClass = {astro-ph.HE},
       adsurl = {https://ui.adsabs.harvard.edu/abs/2019ApJ...887L..24M}
}

@ARTICLE{2021ApJ...918L..27R,
       author = {{Riley}, Thomas E. and {Watts}, Anna L. and {Ray}, Paul S. and {Bogdanov}, Slavko and {Guillot}, Sebastien and {Morsink}, Sharon M. and {Bilous}, Anna V. and {Arzoumanian}, Zaven and {Choudhury}, Devarshi and {Deneva}, Julia S. and {Gendreau}, Keith C. and {Harding}, Alice K. and {Ho}, Wynn C.~G. and {Lattimer}, James M. and {Loewenstein}, Michael and {Ludlam}, Renee M. and {Markwardt}, Craig B. and {Okajima}, Takashi and {Prescod-Weinstein}, Chanda and {Remillard}, Ronald A. and {Wolff}, Michael T. and {Fonseca}, Emmanuel and {Cromartie}, H. Thankful and {Kerr}, Matthew and {Pennucci}, Timothy T. and {Parthasarathy}, Aditya and {Ransom}, Scott and {Stairs}, Ingrid and {Guillemot}, Lucas and {Cognard}, Ismael},
        title = "{A NICER View of the Massive Pulsar PSR J0740+6620 Informed by Radio Timing and XMM-Newton Spectroscopy}",
      journal = {\apjl},
         year = 2021,
        month = sep,
       volume = {918},
       number = {2},
          eid = {L27},
        pages = {L27},
          doi = {10.3847/2041-8213/ac0a81},
archivePrefix = {arXiv},
       eprint = {2105.06980},
 primaryClass = {astro-ph.HE},
       adsurl = {https://ui.adsabs.harvard.edu/abs/2021ApJ...918L..27R}
}

@ARTICLE{2021ApJ...918L..28M,
       author = {{Miller}, M.~C. and {Lamb}, F.~K. and {Dittmann}, A.~J. and {Bogdanov}, S. and {Arzoumanian}, Z. and {Gendreau}, K.~C. and {Guillot}, S. and {Ho}, W.~C.~G. and {Lattimer}, J.~M. and {Loewenstein}, M. and {Morsink}, S.~M. and {Ray}, P.~S. and {Wolff}, M.~T. and {Baker}, C.~L. and {Cazeau}, T. and {Manthripragada}, S. and {Markwardt}, C.~B. and {Okajima}, T. and {Pollard}, S. and {Cognard}, I. and {Cromartie}, H.~T. and {Fonseca}, E. and {Guillemot}, L. and {Kerr}, M. and {Parthasarathy}, A. and {Pennucci}, T.~T. and {Ransom}, S. and {Stairs}, I.},
        title = "{The Radius of PSR J0740+6620 from NICER and XMM-Newton Data}",
      journal = {\apjl},
         year = 2021,
        month = sep,
       volume = {918},
       number = {2},
          eid = {L28},
        pages = {L28},
          doi = {10.3847/2041-8213/ac089b},
archivePrefix = {arXiv},
       eprint = {2105.06979},
 primaryClass = {astro-ph.HE},
       adsurl = {https://ui.adsabs.harvard.edu/abs/2021ApJ...918L..28M}
}

@ARTICLE{2014PhRvC..89f5802C,
       author = {{Constantinou}, Constantinos and {Muccioli}, Brian and {Prakash}, Madappa and {Lattimer}, James M.},
        title = "{Thermal properties of supernova matter: The bulk homogeneous phase}",
      journal = {\prc},
         year = 2014,
        month = jun,
       volume = {89},
       number = {6},
          eid = {065802},
        pages = {065802},
          doi = {10.1103/PhysRevC.89.065802},
archivePrefix = {arXiv},
       eprint = {1402.6348},
 primaryClass = {astro-ph.SR},
       adsurl = {https://ui.adsabs.harvard.edu/abs/2014PhRvC..89f5802C}
}

@ARTICLE{2021EPJA...57..329R,
       author = {{Raduta}, Adriana R. and {Nacu}, Flavia and {Oertel}, Micaela},
        title = "{Equations of state for hot neutron stars}",
      journal = {European Physical Journal A},
         year = 2021,
        month = dec,
       volume = {57},
       number = {12},
          eid = {329},
        pages = {329},
          doi = {10.1140/epja/s10050-021-00628-z},
archivePrefix = {arXiv},
       eprint = {2109.00251},
 primaryClass = {nucl-th},
       adsurl = {https://ui.adsabs.harvard.edu/abs/2021EPJA...57..329R}
}

@ARTICLE{2021PhRvC.103e5806K,
       author = {{Keller}, J. and {Wellenhofer}, C. and {Hebeler}, K. and {Schwenk}, A.},
        title = "{Neutron matter at finite temperature based on chiral effective field theory interactions}",
      journal = {\prc},
         year = 2021,
        month = may,
       volume = {103},
       number = {5},
          eid = {055806},
        pages = {055806},
          doi = {10.1103/PhysRevC.103.055806},
archivePrefix = {arXiv},
       eprint = {2011.05855},
 primaryClass = {nucl-th},
       adsurl = {https://ui.adsabs.harvard.edu/abs/2021PhRvC.103e5806K}
}

@ARTICLE{2023ApJ...952L..36F,
       author = {{Fields}, Jacob and {Prakash}, Aviral and {Breschi}, Matteo and {Radice}, David and {Bernuzzi}, Sebastiano and {Schneider}, Andr{\'e} da Silva},
        title = "{Thermal Effects in Binary Neutron Star Mergers}",
      journal = {\apjl},
         year = 2023,
        month = aug,
       volume = {952},
       number = {2},
          eid = {L36},
        pages = {L36},
          doi = {10.3847/2041-8213/ace5b2},
archivePrefix = {arXiv},
       eprint = {2302.11359},
 primaryClass = {astro-ph.HE},
       adsurl = {https://ui.adsabs.harvard.edu/abs/2023ApJ...952L..36F}
}

@ARTICLE{2025PhRvD.112l3008G,
       author = {{Gieg}, Henrique and {Ujevic}, Maximiliano and {Sedrakian}, Armen and {Dietrich}, Tim},
        title = "{Simulating binary neutron star mergers with finite-temperature equations of state: The influences of the slope of the symmetry energy and artificial heating}",
      journal = {\prd},
         year = 2025,
        month = dec,
       volume = {112},
       number = {12},
          eid = {123008},
        pages = {123008},
          doi = {10.1103/k4mp-ksxy},
archivePrefix = {arXiv},
       eprint = {2507.23684},
 primaryClass = {gr-qc},
       adsurl = {https://ui.adsabs.harvard.edu/abs/2025PhRvD.112l3008G}
}

@ARTICLE{2010PhRvD..82h4043B,
       author = {{Bauswein}, A. and {Janka}, H.-T. and {Oechslin}, R.},
        title = "{Testing approximations of thermal effects in neutron star merger simulations}",
      journal = {\prd},
         year = 2010,
        month = oct,
       volume = {82},
       number = {8},
          eid = {084043},
        pages = {084043},
          doi = {10.1103/PhysRevD.82.084043},
archivePrefix = {arXiv},
       eprint = {1006.3315},
 primaryClass = {astro-ph.SR},
       adsurl = {https://ui.adsabs.harvard.edu/abs/2010PhRvD..82h4043B}
}

@ARTICLE{2020JHEAp..28...19L,
       author = {{Li}, A. and {Zhu}, Z.-Y. and {Zhou}, E.-P. and {Dong}, J.-M. and {Hu}, J.-N. and {Xia}, C.-J.},
        title = "{Neutron star equation of state: Quark mean-field (QMF) modeling and applications}",
      journal = {Journal of High Energy Astrophysics},
         year = 2020,
        month = nov,
       volume = {28},
        pages = {19-46},
          doi = {10.1016/j.jheap.2020.07.001},
archivePrefix = {arXiv},
       eprint = {2007.05116},
 primaryClass = {nucl-th},
       adsurl = {https://ui.adsabs.harvard.edu/abs/2020JHEAp..28...19L}
}

@ARTICLE{2023PhRvC.108b5809Z,
       author = {{Zhu}, Zhenyu and {Li}, Ang and {Hu}, Jinniu and {Shen}, Hong},
        title = "{Equation of state of nuclear matter and neutron stars: Quark mean-field model versus relativistic mean-field model}",
      journal = {\prc},
         year = 2023,
        month = aug,
       volume = {108},
       number = {2},
          eid = {025809},
        pages = {025809},
          doi = {10.1103/PhysRevC.108.025809},
archivePrefix = {arXiv},
       eprint = {2305.16058},
 primaryClass = {nucl-th},
       adsurl = {https://ui.adsabs.harvard.edu/abs/2023PhRvC.108b5809Z}
}

@ARTICLE{2025PhRvD.112d3018T,
       author = {{Tu}, Zhonghao and {Sun}, Xiangdong and {Han}, Shuochong and {Miao}, Zhiqiang and {Li}, Ang},
        title = "{Unified quark mean-field equation of state for neutron star matter: Static and dynamic properties}",
      journal = {\prd},
         year = 2025,
        month = aug,
       volume = {112},
       number = {4},
          eid = {043018},
        pages = {043018},
          doi = {10.1103/bmb3-ktz6},
archivePrefix = {arXiv},
       eprint = {2505.00539},
 primaryClass = {nucl-th},
       adsurl = {https://ui.adsabs.harvard.edu/abs/2025PhRvD.112d3018T}
}

@ARTICLE{2018ApJ...862...98Z,
       author = {{Zhu}, Zhen-Yu and {Zhou}, En-Ping and {Li}, Ang},
        title = "{Neutron Star Equation of State from the Quark Level in Light of GW170817}",
      journal = {\apj},
         year = 2018,
        month = aug,
       volume = {862},
       number = {2},
          eid = {98},
        pages = {98},
          doi = {10.3847/1538-4357/aacc28},
archivePrefix = {arXiv},
       eprint = {1802.05510},
 primaryClass = {nucl-th},
       adsurl = {https://ui.adsabs.harvard.edu/abs/2018ApJ...862...98Z}
}

@ARTICLE{2019PhRvC..99b5804Z,
       author = {{Zhu}, Zhen-Yu and {Li}, Ang and {Hu}, Jin-Niu and {Shen}, Hong},
        title = "{Quark mean-field model for nuclear matter with or without bag}",
      journal = {\prc},
         year = 2019,
        month = feb,
       volume = {99},
       number = {2},
          eid = {025804},
        pages = {025804},
          doi = {10.1103/PhysRevC.99.025804},
archivePrefix = {arXiv},
       eprint = {1805.04678},
 primaryClass = {nucl-th},
       adsurl = {https://ui.adsabs.harvard.edu/abs/2019PhRvC..99b5804Z}
}

@ARTICLE{2018PhRvC..98f5806A,
       author = {{Alford}, Mark G. and {Harris}, Steven P.},
        title = "{{\ensuremath{\beta}} equilibrium in neutron-star mergers}",
      journal = {\prc},
         year = 2018,
        month = dec,
       volume = {98},
       number = {6},
          eid = {065806},
        pages = {065806},
          doi = {10.1103/PhysRevC.98.065806},
archivePrefix = {arXiv},
       eprint = {1803.00662},
 primaryClass = {nucl-th},
       adsurl = {https://ui.adsabs.harvard.edu/abs/2018PhRvC..98f5806A}
}

@ARTICLE{1999PhRvC..59..682B,
       author = {{Baldo}, M. and {Ferreira}, L.~S.},
        title = "{Nuclear liquid-gas phase transition}",
      journal = {\prc},
         year = 1999,
        month = feb,
       volume = {59},
       number = {2},
        pages = {682-703},
          doi = {10.1103/PhysRevC.59.682},
       adsurl = {https://ui.adsabs.harvard.edu/abs/1999PhRvC..59..682B}
}

@ARTICLE{2012PhRvC..85a4318A,
       author = {{Aguirre}, R.},
        title = "{Specific heat capacity in the low-density regime of asymmetric nuclear matter}",
      journal = {\prc},
         year = 2012,
        month = jan,
       volume = {85},
       number = {1},
          eid = {014318},
        pages = {014318},
          doi = {10.1103/PhysRevC.85.014318},
       adsurl = {https://ui.adsabs.harvard.edu/abs/2012PhRvC..85a4318A}
}

@ARTICLE{2001PhRvC..63e4604S,
       author = {{Sil}, Tapas and {Agrawal}, B.~K. and {de}, J.~N. and {Samaddar}, S.~K.},
        title = "{Liquid-gas phase transition in nuclei in the relativistic Thomas-Fermi theory}",
      journal = {\prc},
         year = 2001,
        month = may,
       volume = {63},
       number = {5},
          eid = {054604},
        pages = {054604},
          doi = {10.1103/PhysRevC.63.054604},
archivePrefix = {arXiv},
       eprint = {nucl-th/0012061},
 primaryClass = {nucl-th},
       adsurl = {https://ui.adsabs.harvard.edu/abs/2001PhRvC..63e4604S}
}

@ARTICLE{2007PhRvL..99l1102O,
       author = {{Oechslin}, R. and {Janka}, H.-T.},
        title = "{Gravitational Waves from Relativistic Neutron-Star Mergers with Microphysical Equations of State}",
      journal = {\prl},
         year = 2007,
        month = sep,
       volume = {99},
       number = {12},
          eid = {121102},
        pages = {121102},
          doi = {10.1103/PhysRevLett.99.121102},
archivePrefix = {arXiv},
       eprint = {astro-ph/0702228},
 primaryClass = {astro-ph},
       adsurl = {https://ui.adsabs.harvard.edu/abs/2007PhRvL..99l1102O}
}

@ARTICLE{2024PhRvD.110d3002R,
       author = {{Raithel}, Carolyn A. and {Paschalidis}, Vasileios},
        title = "{Detectability of finite-temperature effects from neutron star mergers with next-generation gravitational wave detectors}",
      journal = {\prd},
         year = 2024,
        month = aug,
       volume = {110},
       number = {4},
          eid = {043002},
        pages = {043002},
          doi = {10.1103/PhysRevD.110.043002},
archivePrefix = {arXiv},
       eprint = {2312.14046},
 primaryClass = {astro-ph.HE},
       adsurl = {https://ui.adsabs.harvard.edu/abs/2024PhRvD.110d3002R}
}

@ARTICLE{2022MNRAS.516.4792R,
       author = {{Raithel}, Carolyn A. and {Espino}, Pedro and {Paschalidis}, Vasileios},
        title = "{Finite-temperature effects in dynamical spacetime binary neutron star merger simulations: validation of the parametric approach}",
      journal = {\mnras},
         year = 2022,
        month = nov,
       volume = {516},
       number = {4},
        pages = {4792-4804},
          doi = {10.1093/mnras/stac2450},
archivePrefix = {arXiv},
       eprint = {2206.14838},
 primaryClass = {astro-ph.HE},
       adsurl = {https://ui.adsabs.harvard.edu/abs/2022MNRAS.516.4792R}
}

@ARTICLE{2021PhRvD.104f3016R,
       author = {{Raithel}, Carolyn A. and {Paschalidis}, Vasileios and {{\"O}zel}, Feryal},
        title = "{Realistic finite-temperature effects in neutron star merger simulations}",
      journal = {\prd},
         year = 2021,
        month = sep,
       volume = {104},
       number = {6},
          eid = {063016},
        pages = {063016},
          doi = {10.1103/PhysRevD.104.063016},
archivePrefix = {arXiv},
       eprint = {2104.07226},
 primaryClass = {astro-ph.HE},
       adsurl = {https://ui.adsabs.harvard.edu/abs/2021PhRvD.104f3016R}
}

@ARTICLE{2026PhRvD.113d3060R,
       author = {{Raithel}, Carolyn A. and {Paschalidis}, Vasileios},
        title = "{Impact of thermal effects on prompt-collapse binary neutron star mergers}",
      journal = {\prd},
         year = 2026,
        month = feb,
       volume = {113},
       number = {4},
          eid = {043060},
        pages = {043060},
          doi = {10.1103/rs9d-dhkm},
archivePrefix = {arXiv},
       eprint = {2508.09333},
 primaryClass = {astro-ph.HE},
       adsurl = {https://ui.adsabs.harvard.edu/abs/2026PhRvD.113d3060R}
}

@ARTICLE{2024arXiv240419756L,
       author = {{Liu}, Ziming and {Wang}, Yixuan and {Vaidya}, Sachin and {Ruehle}, Fabian and {Halverson}, James and {Solja{\v{c}}i{\'c}}, Marin and {Hou}, Thomas Y. and {Tegmark}, Max},
        title = "{KAN: Kolmogorov-Arnold Networks}",
      journal = {arXiv e-prints},
         year = 2024,
        month = apr,
          eid = {arXiv:2404.19756},
        pages = {arXiv:2404.19756},
          doi = {10.48550/arXiv.2404.19756},
archivePrefix = {arXiv},
       eprint = {2404.19756},
 primaryClass = {cs.LG},
       adsurl = {https://ui.adsabs.harvard.edu/abs/2024arXiv240419756L}
}

@ARTICLE{2025PhRvX..15d1051L,
       author = {{Liu}, Ziming and {Tegmark}, Max and {Ma}, Pingchuan and {Matusik}, Wojciech and {Wang}, Yixuan},
        title = "{Kolmogorov-Arnold Networks Meet Science}",
      journal = {Physical Review X},
         year = 2025,
        month = oct,
       volume = {15},
       number = {4},
          eid = {041051},
        pages = {041051},
          doi = {10.1103/4t7t-v19l},
archivePrefix = {arXiv},
       eprint = {2408.10205},
 primaryClass = {cs.LG},
       adsurl = {https://ui.adsabs.harvard.edu/abs/2025PhRvX..15d1051L}
}

@ARTICLE{2025PhRvC.111b4316L,
       author = {{Liu}, Hao and {Lei}, Jin and {Ren}, Zhongzhou},
        title = "{Kolmogorov-Arnold networks in nuclear binding energy prediction}",
      journal = {\prc},
         year = 2025,
        month = feb,
       volume = {111},
       number = {2},
          eid = {024316},
        pages = {024316},
          doi = {10.1103/PhysRevC.111.024316},
archivePrefix = {arXiv},
       eprint = {2407.20737},
 primaryClass = {nucl-th},
       adsurl = {https://ui.adsabs.harvard.edu/abs/2025PhRvC.111b4316L}
}

@ARTICLE{2025PhRvC.112e4002B,
       author = {{Bedaque}, Paulo F. and {Cigliano}, Jacob and {Kumar}, Hersh and {Paul}, Srijit and {Rajawat}, Suryansh},
        title = "{Kolmogorov-Arnold wavefunctions}",
      journal = {\prc},
         year = 2025,
        month = nov,
       volume = {112},
       number = {5},
          eid = {054002},
        pages = {054002},
          doi = {10.1103/zj4l-h6nv},
archivePrefix = {arXiv},
       eprint = {2506.02171},
 primaryClass = {nucl-th},
       adsurl = {https://ui.adsabs.harvard.edu/abs/2025PhRvC.112e4002B}
}

@ARTICLE{2026NuPhA106823327S,
       author = {{Shree}, S. Madhumitha and {Balasubramaniam}, M.},
        title = "{Machine learning and symbolic regression-based modeling of {\ensuremath{\alpha}}-decay half-lives for superheavy nuclei}",
      journal = {\nphysa},
         year = 2026,
        month = apr,
       volume = {1068},
          eid = {123327},
        pages = {123327},
          doi = {10.1016/j.nuclphysa.2026.123327},
       adsurl = {https://ui.adsabs.harvard.edu/abs/2026NuPhA106823327S}
}

@ARTICLE{2025PhRvD.111h3009G,
       author = {{Grippa}, Francesco and {Prakash}, Aviral and {Logoteta}, Domenico and {Radice}, David and {Bombaci}, Ignazio},
        title = "{General relativistic hydrodynamic simulations of binary strange star mergers}",
      journal = {\prd},
         year = 2025,
        month = apr,
       volume = {111},
       number = {8},
          eid = {083009},
        pages = {083009},
          doi = {10.1103/PhysRevD.111.083009},
archivePrefix = {arXiv},
       eprint = {2407.11143},
 primaryClass = {astro-ph.HE},
       adsurl = {https://ui.adsabs.harvard.edu/abs/2025PhRvD.111h3009G}
}

@ARTICLE{2025PhRvD.112j3003Z,
       author = {{Zhu}, Zhenyu and {Zha}, Shuai and {Han}, Sophia},
        title = "{Three-dimensional equation of state extension of quark matter in the Fermi-liquid theory}",
      journal = {\prd},
         year = 2025,
        month = nov,
       volume = {112},
       number = {10},
          eid = {103003},
        pages = {103003},
          doi = {10.1103/f8tf-3dn5},
archivePrefix = {arXiv},
       eprint = {2506.17569},
 primaryClass = {astro-ph.HE},
       adsurl = {https://ui.adsabs.harvard.edu/abs/2025PhRvD.112j3003Z}
}

@ARTICLE{2019PhRvC.100e5802S,
       author = {{Schneider}, A.~S. and {Roberts}, L.~F. and {Ott}, C.~D. and {O'Connor}, E.},
        title = "{Equation of state effects in the core collapse of a 20 -M$_{{\ensuremath{\odot}}}$ star}",
      journal = {\prc},
         year = 2019,
        month = nov,
       volume = {100},
       number = {5},
          eid = {055802},
        pages = {055802},
          doi = {10.1103/PhysRevC.100.055802},
archivePrefix = {arXiv},
       eprint = {1906.02009},
 primaryClass = {astro-ph.HE},
       adsurl = {https://ui.adsabs.harvard.edu/abs/2019PhRvC.100e5802S}
}

@ARTICLE{2020PhRvL.124i2701Y,
       author = {{Yasin}, H. and {Sch{\"a}fer}, S. and {Arcones}, A. and {Schwenk}, A.},
        title = "{Equation of State Effects in Core-Collapse Supernovae}",
      journal = {\prl},
         year = 2020,
        month = mar,
       volume = {124},
       number = {9},
          eid = {092701},
        pages = {092701},
          doi = {10.1103/PhysRevLett.124.092701},
archivePrefix = {arXiv},
       eprint = {1812.02002},
 primaryClass = {nucl-th},
       adsurl = {https://ui.adsabs.harvard.edu/abs/2020PhRvL.124i2701Y}
}

@ARTICLE{2020ApJ...894....4D,
       author = {{Schneider}, Andr{\'e} da Silva and {O'Connor}, Evan and {Granqvist}, Elvira and {Betranhandy}, Aurore and {Couch}, Sean M.},
        title = "{Equation of State and Progenitor Dependence of Stellar-mass Black Hole Formation}",
      journal = {\apj},
         year = 2020,
        month = may,
       volume = {894},
       number = {1},
          eid = {4},
        pages = {4},
          doi = {10.3847/1538-4357/ab8308},
archivePrefix = {arXiv},
       eprint = {2001.10434},
 primaryClass = {astro-ph.HE},
       adsurl = {https://ui.adsabs.harvard.edu/abs/2020ApJ...894....4D}
}

@ARTICLE{2025ApJ...980...53E,
       author = {{Eggenberger Andersen}, Oliver and {O'Connor}, Evan and {Andresen}, Haakon and {da Silva Schneider}, Andr{\'e} and {Couch}, Sean M.},
        title = "{Black Hole Supernovae, Their Equation of State Dependence, and Ejecta Composition}",
      journal = {\apj},
         year = 2025,
        month = feb,
       volume = {980},
       number = {1},
          eid = {53},
        pages = {53},
          doi = {10.3847/1538-4357/ada899},
archivePrefix = {arXiv},
       eprint = {2411.11969},
 primaryClass = {astro-ph.HE},
       adsurl = {https://ui.adsabs.harvard.edu/abs/2025ApJ...980...53E}
}

@ARTICLE{2026arXiv260906374Z,
       author = {{Zhu}, Zhenyu and {Chabanov}, Michail and {O'Shaughnessy}, Richard},
        title = "{Detectability of bulk viscosity effects on post-merger gravitational wave signals from binary neutron star mergers}",
      journal = {arXiv e-prints},
         year = 2026,
        month = sep,
          eid = {arXiv:2609.06374},
        pages = {arXiv:2609.06374},
          doi = {10.48550/arXiv.2609.06374},
archivePrefix = {arXiv},
       eprint = {2609.06374},
 primaryClass = {astro-ph.HE},
       adsurl = {https://ui.adsabs.harvard.edu/abs/2026arXiv260906374Z}
}

@ARTICLE{2018PhRvL.120d1101A,
       author = {{Alford}, Mark G. and {Bovard}, Luke and {Hanauske}, Matthias and {Rezzolla}, Luciano and {Schwenzer}, Kai},
        title = "{Viscous Dissipation and Heat Conduction in Binary Neutron-Star Mergers}",
      journal = {\prl},
         year = 2018,
        month = jan,
       volume = {120},
       number = {4},
          eid = {041101},
        pages = {041101},
          doi = {10.1103/PhysRevLett.120.041101},
archivePrefix = {arXiv},
       eprint = {1707.09475},
 primaryClass = {gr-qc},
       adsurl = {https://ui.adsabs.harvard.edu/abs/2018PhRvL.120d1101A}
}

@ARTICLE{2025PhRvD.111d4074C,
       author = {{Chabanov}, Michail and {Rezzolla}, Luciano},
        title = "{Numerical modeling of bulk viscosity in neutron stars}",
      journal = {\prd},
         year = 2025,
        month = feb,
       volume = {111},
       number = {4},
          eid = {044074},
        pages = {044074},
          doi = {10.1103/PhysRevD.111.044074},
       adsurl = {https://ui.adsabs.harvard.edu/abs/2025PhRvD.111d4074C}
}

@ARTICLE{2025PhRvL.134g1402C,
       author = {{Chabanov}, Michail and {Rezzolla}, Luciano},
        title = "{Impact of Bulk Viscosity on the Postmerger Gravitational-Wave Signal from Merging Neutron Stars}",
      journal = {\prl},
         year = 2025,
        month = feb,
       volume = {134},
       number = {7},
          eid = {071402},
        pages = {071402},
          doi = {10.1103/PhysRevLett.134.071402},
archivePrefix = {arXiv},
       eprint = {2307.10464},
 primaryClass = {gr-qc},
       adsurl = {https://ui.adsabs.harvard.edu/abs/2025PhRvL.134g1402C}
}

@ARTICLE{2022MNRAS.509.1096M,
       author = {{Most}, Elias R. and {Harris}, Steven P. and {Plumberg}, Christopher and {Alford}, Mark G. and {Noronha}, Jorge and {Noronha-Hostler}, Jacquelyn and {Pretorius}, Frans and {Witek}, Helvi and {Yunes}, Nicol{\'a}s},
        title = "{Projecting the likely importance of weak-interaction-driven bulk viscosity in neutron star mergers}",
      journal = {\mnras},
         year = 2022,
        month = jan,
       volume = {509},
       number = {1},
        pages = {1096-1108},
          doi = {10.1093/mnras/stab2793},
archivePrefix = {arXiv},
       eprint = {2107.05094},
 primaryClass = {astro-ph.HE},
       adsurl = {https://ui.adsabs.harvard.edu/abs/2022MNRAS.509.1096M}
}

@ARTICLE{2024ApJ...967L..14M,
       author = {{Most}, Elias R. and {Haber}, Alexander and {Harris}, Steven P. and {Zhang}, Ziyuan and {Alford}, Mark G. and {Noronha}, Jorge},
        title = "{Emergence of Microphysical Bulk Viscosity in Binary Neutron Star Postmerger Dynamics}",
      journal = {\apjl},
         year = 2024,
        month = may,
       volume = {967},
       number = {1},
          eid = {L14},
        pages = {L14},
          doi = {10.3847/2041-8213/ad454f},
archivePrefix = {arXiv},
       eprint = {2207.00442},
 primaryClass = {astro-ph.HE},
       adsurl = {https://ui.adsabs.harvard.edu/abs/2024ApJ...967L..14M}
}

@ARTICLE{2024PhRvL.132u1001E,
       author = {{Espino}, Pedro Luis and {Hammond}, Peter and {Radice}, David and {Bernuzzi}, Sebastiano and {Gamba}, Rossella and {Zappa}, Francesco and {Micchi}, Lu{\'\i}s Felipe Longo and {Perego}, Albino},
        title = "{Neutrino Trapping and Out-of-Equilibrium Effects in Binary Neutron-Star Merger Remnants}",
      journal = {\prl},
         year = 2024,
        month = may,
       volume = {132},
       number = {21},
          eid = {211001},
        pages = {211001},
          doi = {10.1103/PhysRevLett.132.211001},
archivePrefix = {arXiv},
       eprint = {2311.00031},
 primaryClass = {astro-ph.HE},
       adsurl = {https://ui.adsabs.harvard.edu/abs/2024PhRvL.132u1001E}
}

@ARTICLE{2026arXiv260323904H,
       author = {{Hern{\'a}ndez}, Jos{\'e} Luis and {Manuel}, Cristina and {Tolos}, Laura},
        title = "{Impact of muons on the bulk viscosity of neutron star matter metamodels}",
      journal = {arXiv e-prints},
         year = 2026,
        month = mar,
          eid = {arXiv:2603.23904},
        pages = {arXiv:2603.23904},
          doi = {10.48550/arXiv.2603.23904},
archivePrefix = {arXiv},
       eprint = {2603.23904},
 primaryClass = {hep-ph},
       adsurl = {https://ui.adsabs.harvard.edu/abs/2026arXiv260323904H}
}

@ARTICLE{2023PhRvD.108j3037R,
       author = {{Ripley}, Justin L. and {Abhishek Hegade K.}, R. and {Yunes}, Nicol{\'a}s},
        title = "{Probing internal dissipative processes of neutron stars with gravitational waves during the inspiral of neutron star binaries}",
      journal = {\prd},
         year = 2023,
        month = nov,
       volume = {108},
       number = {10},
          eid = {103037},
        pages = {103037},
          doi = {10.1103/PhysRevD.108.103037},
archivePrefix = {arXiv},
       eprint = {2306.15633},
 primaryClass = {gr-qc},
       adsurl = {https://ui.adsabs.harvard.edu/abs/2023PhRvD.108j3037R}
}

@ARTICLE{2024NatAs...8.1277R,
       author = {{Ripley}, Justin L. and {Hegade K R}, Abhishek and {Chandramouli}, Rohit S. and {Yunes}, Nicol{\'a}s},
        title = "{A constraint on the dissipative tidal deformability of neutron stars}",
      journal = {Nature Astronomy},
         year = 2024,
        month = oct,
       volume = {8},
       number = {10},
        pages = {1277-1283},
          doi = {10.1038/s41550-024-02323-7},
archivePrefix = {arXiv},
       eprint = {2312.11659},
 primaryClass = {gr-qc},
       adsurl = {https://ui.adsabs.harvard.edu/abs/2024NatAs...8.1277R}
}

\end{document}